\documentclass{article}
\usepackage{arxiv}
\usepackage[utf8]{inputenc}
\usepackage[T1]{fontenc}
\usepackage{mathptmx}

\usepackage[american]{babel}
\usepackage{csquotes}

\usepackage[style=numeric, sorting=none, backend=biber]{biblatex}
\usepackage{amsmath}
\usepackage{amssymb}
\usepackage{amsfonts}
\usepackage{nicefrac}

\usepackage{booktabs}
\usepackage{array}
\usepackage{hhline}
\usepackage{makecell}
\usepackage{multirow}
\usepackage{tabularx}

\usepackage{graphicx}
\usepackage{float}
\usepackage{caption}
\usepackage{geometry}
\usepackage{pdflscape}

\usepackage{setspace}
\AtBeginEnvironment{tabular}{\onehalfspacing}
\AtBeginEnvironment{tablenotes}{\doublespacing}
\usepackage{hyperref}
\usepackage{url}
\usepackage{doi}

\usepackage{microtype}
\usepackage{cleveref}

\hypersetup{
  pdftitle={Measuring successful cooperation in human-AI teamwork},
  pdfsubject={human-computer interaction},
  pdfauthor={Christiane Attig et al.},
  pdfkeywords={human-AI teaming, human-AI interaction, cooperation, user experience, scale development},
}

\renewcommand{\shorttitle}{Measuring Successful Human-AI Teamwork}
\renewcommand{\undertitle}{Preprint submitted for Publication}
\date{}
\renewcommand{\headeright}{Attig et al.}

\title{Measuring Successful Cooperation in Human-AI Teamwork: Development and
Validation of the Perceived Cooperativity and Teaming Perception Scales}

\usepackage{authblk}

\newbox{\orcid}\sbox{\orcid}{\includegraphics[scale=0.06]{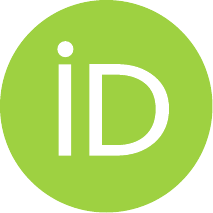}}

\author[1,2]{%
  \href{https://orcid.org/0000-0002-6280-2530}{\usebox{\orcid}\hspace{1mm}Christiane Attig}%
}
\author[1]{%
  \href{https://orcid.org/0000-0002-6574-3820}{\usebox{\orcid}\hspace{1mm}Christiane Wiebel-Herboth}%
}
\author[1]{%
  \href{https://orcid.org/0000-0002-7105-5207}{\usebox{\orcid}\hspace{1mm}Patricia Wollstadt}%
}
\author[2]{%
  \href{https://orcid.org/0000-0001-7685-1598}{\usebox{\orcid}\hspace{1mm}Tim Schrills}%
}
\author[2]{%
  \href{https://orcid.org/0000-0003-4458-8443}{\usebox{\orcid}\hspace{1mm}Mourad Zoubir}%
}
\author[2]{%
  \href{https://orcid.org/0000-0002-7211-3771}{\usebox{\orcid}\hspace{1mm}Thomas Franke}%
}
\affil[1]{Honda Research Institute Europe GmbH, Offenbach am Main, Germany}
\affil[2]{Institute of Human-Centered Interactive Systems, University of Lübeck, Lübeck, Germany}
\begin{document}

\maketitle

\vspace{-2em}
\begin{abstract}
As human–AI cooperation becomes increasingly prevalent, reliable instruments for assessing the subjective quality of cooperative human–AI interaction are needed. We introduce two theoretically grounded scales: the Perceived Cooperativity Scale (PCS), grounded in joint activity theory, and the Teaming Perception Scale (TPS), grounded in evolutionary cooperation theory. The PCS captures an agent's perceived cooperative capability and practice within a single interaction sequence; the TPS captures the emergent sense of teaming arising from mutual contribution and support. Both scales were adapted for human–human cooperation to enable cross-agent comparisons. Across three studies ($N = 409$) encompassing a cooperative card game, LLM interaction, and a decision-support system, analyses of dimensionality, reliability, and validity indicated that both scales successfully differentiated between cooperation partners of varying cooperative quality and showed construct validity in line with expectations. The scales provide a basis for empirical investigation and system evaluation across a wide range of human–AI cooperation contexts.
\end{abstract}

\section{Introduction}
Over the past few years, developments in artificial intelligence (AI) have been profoundly transformative. What has once been the realm of science fiction---conversing in natural language with artificial agents that remember prior interactions, assist with a wide range of tasks, and can be perceived as team members, friends, or even relationship partners---has become part of daily life for millions of people \parencite{WOLF2024102821}. This shift is largely driven by Large Language Models (LLMs) based on generative pre-trained transformers (GPTs; \cite{akhtarUnveilingEvolutionGenerative2024}). In parallel, other machine and deep learning systems have advanced, for instance, in visual pattern recognition (e.g., cancer detection in medical imaging; \cite{javedDeepLearningLungs2024}, lane detection in autonomous vehicle control; \cite{zakaria2023lanedetection}, threat detection in critical infrastructure; \cite{budzysDeepLearningbasedAuthentication2024}) and prediction (e.g., of pandemic diseases; \cite{ajagbeDeepLearningTechniques2024}, market trends; \cite{al-khasawnehStockMarketTrend2025}, or energy consumption; \cite{zhangReviewMachineLearning2024}), broadening AI’s reach across sectors such as private life, healthcare, transportation, and industry.

Given that AI systems are integrated in an increasing number of applications and become part of users’ daily routines, understanding the psychological dynamics that arise during human-AI interaction is crucial to enable AI experiences that have a positive impact on individual user satisfaction as well as societal flourishing \parencite{attigMoreTaskPerformance2024}. The importance of this understanding is stressed by first findings indicating potentially adverse implications of AI use on individual, group, and societal levels. For instance, on the individual level, interacting with AI at work has the potential to heighten feelings of loneliness \parencite{tangNoPersonIsland2023}, boredom, and demotivation \parencite{wuHumangenerativeAICollaboration2025a}. On the group level, integrating an AI into a hybrid team might lead to deteriorating team cognition \parencite{mcneeseWhoWhatMy2021,sidjiHumanAICollaborationCooperative2024}. On broader social structures, AI interaction might cause negative spillover effects, such as undermining social norms of reciprocity in traffic settings \parencite{shiradoEmergenceCollapseReciprocity2023} or of politeness in communicative situations \parencite{teyPeopleJudgeOthers2024}. Thus, a core challenge for human factors and engineering psychology is to advance theoretical and methodological foundations that can account for these multi-level effects and guide the development of human-AI interactions that are as effective as they are aligned with human psychology and social responsibility \parencite{attigMoreTaskPerformance2024, schmagerUnderstandingHumanCentredAI, sendhoffCooperativeIntelligenceHumane07}.

Importantly, such individual and societal effects are likely shaped by the depth to which humans and AI systems are \emph{integrated}. Regarding a single task---what we term \emph{synchronic integration}---human-AI coupling can vary from minimal, task-specific assistance to highly interdependent integration of human and AI information processing and behavior \parencite{johnsonTomorrowsHumanMachine2018, oneillHumanAutonomyTeaming2022}. Moreover, when looking beyond a single interaction---what we term \emph{diachronic integration}---as in repeated collaborations and long-term cooperation, the degree of coupling can range from occasional, loosely connected joint actions to enduring partnerships in which humans and AI systems develop shared routines, mutual influence, and a joint history of interaction \parencite{chang2025frominteraction, kirkWhyHumanAI2025, qianCoexperiencingAIEffects2025}. Particularly regarding cooperative tasks in performance-related settings (e.g., clinical decision-making, collaborative problem-solving, education), tighter synchronic and diachronic integration can blur boundaries between human and AI contributions, influencing roles, responsibilities, and the dynamics of teamwork \parencite{basappaMindGapsHow2025, schelbleShouldAITeammates2025}. Thus, hybrid teams with high interdependence of tasks represent a context in which individual, team, and potentially societal dynamics are especially sensitive to disruption by AI, highlighting the urgent need to understand and guide these interactions through human-centered design.

A key prerequisite for understanding the factors that shape effective and socially responsible human-AI cooperation is the availability of context-specific instruments that can reliably capture humans’ subjective perceptions of the cooperative dynamics. While a large variety of questionnaire scales have been developed over the last decade to assess users’ general attitudes towards AI \parencite{steinAttitudesAIMeasurement2024a} and AI literacy \parencite{lintnerSystematicReviewAI2024}, there remains a need for specialized instruments capable of capturing users’ subjective perceptions of specific cooperative human-AI interaction qualities and team dynamics. In the present work, we introduce two theoretically grounded scales designed to reliably assess users’ perceptions of both the cooperative capabilities of artificial agents and the broader dynamics of hybrid teamwork. Moreover, the scales were also adapted for human-human teamwork to enable comparisons of AI and human cooperation partners. Across three studies encompassing different contexts and levels of human-AI integration, we collected data to validate the questionnaires. The resulting scales not only enable rigorous empirical investigation of hybrid teamwork dynamics but also provide practical guidance for designing and evaluating AI systems that can integrate seamlessly into collaborative environments.

\section{Background}
\subsection{Synchronic and diachronic dimensions of human-AI integration}\label{sec:SynchronicDiachronic}
\begin{figure}
    \centering
    \includegraphics[width=0.65\textwidth]{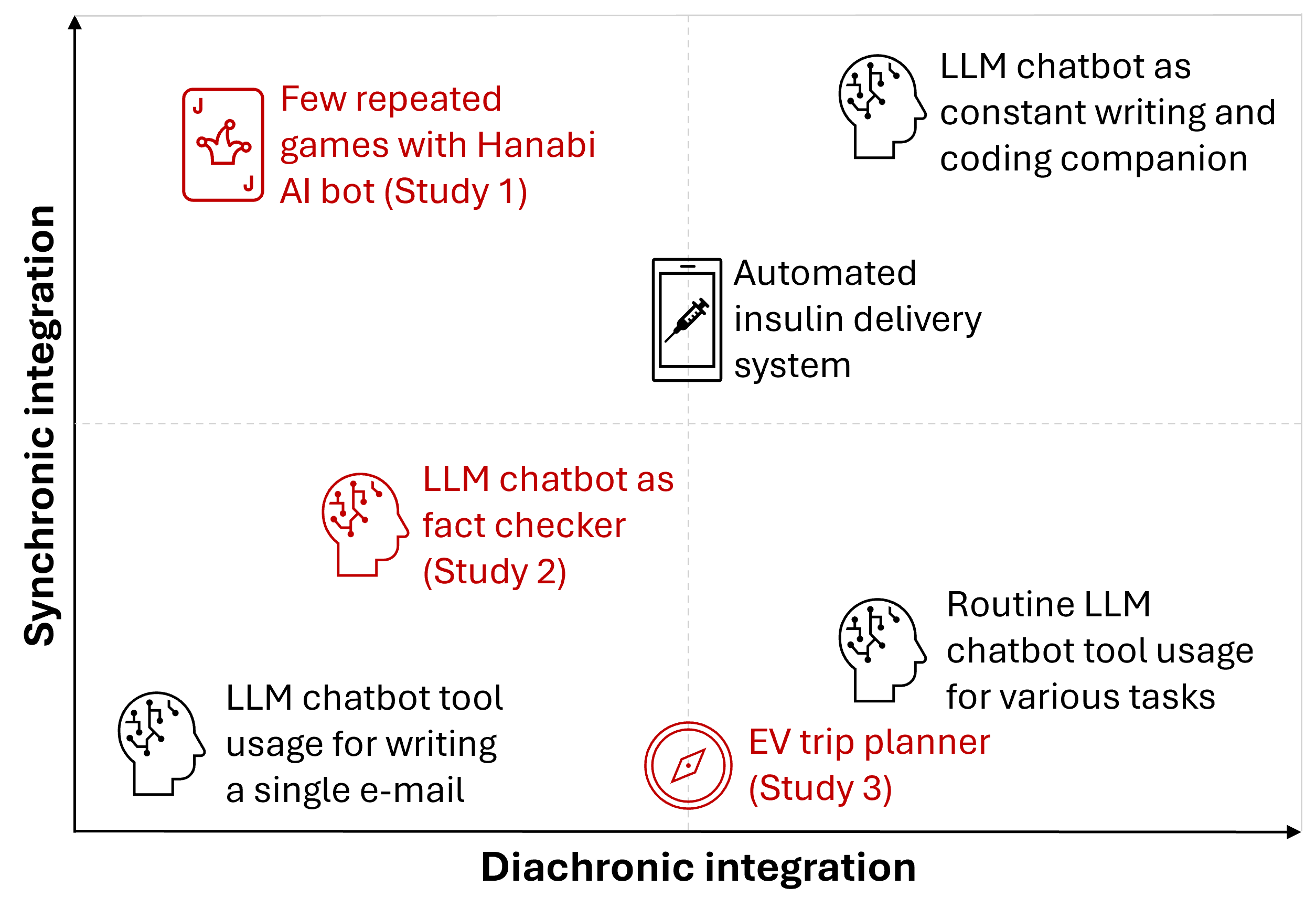}
    \caption {Illustration of two-axis model of synchronic and diachronic dimensions of human-AI integration. Synchronic integration refers to the degree of integration between human and AI information processing and behavior within a single task sequence. Diachronic integration refers to the degree of integration between human and AI information processing across different repeated task sequences.}
    \label{fig:illustration}
\end{figure}

To evaluate the success of human–AI cooperation on a psychological level, it is essential to first clarify the degree of integration between human and AI partners and the durability of their cooperation, since different psychological processes come into play at varying levels of integration and durability. For a systematic understanding of the cooperative structure, we propose a differentiation between a \emph{synchronic} and a \emph{diachronic} dimension of integration (see Figure \ref{fig:illustration}). Borrowed from linguistics where they were introduced by Ferdinand de Saussure, these terms distinguish between phenomena examined at a single point in time (synchronic) and phenomena examined as they unfold and develop over time (diachronic; \cite{buchanan2018dictionary}).

With the term \emph{synchronic integration} we refer to the degree of integration between human and AI information processing and behavior within a \emph{single task sequence}. Single tasks with low interdependency between partners are typically characterized by loose synchronic integration, meaning that the task is mostly completed either by the human or by the AI system. For instance, low synchronic integration occurs when a user submits a simple prompt to an LLM-based chatbot (e.g., ChatGPT, Claude) to draft an email to their doctor and sends the generated text without modification. In contrast, high synchronic integration involves a more iterative and tightly coupled process, where the user provides a more elaborated initial prompt, evaluates the AI’s response, refines the prompt, and repeats this cycle until the output meets their requirements. This illustrates cognitive interdependence \parencite{alqasirRelationalShiftWhy2025a}, where human and AI information processing and action regulation are closely intertwined in shaping the final outcome.

In contrast, with the term \emph{diachronic integration} we refer to the degree of integration between human and AI information processing across different \emph{repeated task sequences}. Loose diachronic integration is characterized by occasional joint actions. For instance, imagine a user who consults an LLM-based chatbot every few months when drafting professional emails, relying on the AI only sporadically and performing most of the remaining work independently. In contrast, deep diachronic integration occurs when a user routinely consults the chatbot across a wide range of tasks over time, for instance, using it briefly for email drafting one day, brainstorming meeting agendas the next, and looking up information for a report the week after. Across these interactions, the user develops stable reliance patterns, learns the AI's capabilities and limitations, and gradually integrates the system as a consistently present entity in their workflow---even though no single interaction may involve deep synchronic coupling. High diachronic integration, however, can also co-occur with high synchronic integration, as when a user collaborates with the chatbot continuously and iteratively across multiple demanding tasks such as manuscript improvement, coding, and data analysis, developing shared routines in which the AI consistently supports planning, drafting, and refining outputs. As another example, consider an automated insulin delivery (AID) system \parencite{schrillsSafeEnvironmentsUnderstand2023}. The system is continuously coupled with the human user at a physical level, while the integration of human and AI information processing is better characterized as episodic and interdependent (arising in specific moments, e.g., when the user checks measured glucose levels, when the system prompts the user for manual input regarding physical exercise). Even though the AID system is used over extended periods of time, the interaction remains confined to a single, physiologically defined task domain. Therefore, despite the continuity of use, diachronic integration remains at a medium level.

In sum, while synchronic integration refers to the level of interdependence within a single task sequence, diachronic integration captures the continuity and stability of cooperation across multiple task sequences over time. Consequently, synchronic and diachronic integration are assumed to be relatively independent dimensions. However, the assumed consequences of human-AI coupling on human information processing (e.g., long-term effects of cognitive offloading, \cite{shukla2025deskilling}) and hybrid team dynamics (e.g., team development, \cite{schelbleShouldAITeammates2025}) are more likely to manifest when both synchronic and diachronic integration are high.

Knowledge about the depth of synchronic and diachronic integration in a given human–AI cooperation process is crucial for (a) identifying the potential psychological dynamics at play, (b) selecting appropriate theoretical frameworks to explain those dynamics, (c) and determining suitable metrics for assessing subjective cooperation success. When diachronic integration is absent or minimal---ranging from single interaction instances to only occasional joint actions---key questions concern the degree of synchronic integration, the extent to which cooperation partners are prepared for this level of interdependence, and the potential consequences that follow. For instance, strong synchronic integration with highly interdependent tasks typically heightens the need for coordination between partners \parencite{klein2005common}. If either partner lacks the necessary capabilities for such coordination, the success of cooperation may be compromised \parencite{attigMoreTaskPerformance2024}. In contrast, when examining repeated cooperative interactions characterized by high diachronic integration, as for instance in hybrid work environments, consequences of integrating AI agents on team dynamics (e.g., team cohesion, \cite{lakhmaniCohesionHumanAutonomy2022}; trust calibration over time, \cite{devisserTheoryLongitudinalTrust2020}) become more relevant. Hence, different theoretical foundations are required to adequately capture the dynamics of human–AI cooperation, with synchronically integrated interactions benefiting from theories of interdependence, while diachronically integrated interactions call more strongly for longitudinal perspectives on cooperation.






\subsection{From joint activity theory to perceived cooperativity}

A fruitful foundation for systematically investigating synchronic human–AI integration is joint activity theory \parencite{klein2005common, bradshaw2017human}, which conceptualizes joint activity as an interdependent process that requires the willingness of all parties to contribute to its successful completion. As it centers on the choreography of single joint activities, the theory is particularly well suited to examining the factors that determine the success of individual cooperative task sequences.

According to joint activity theory, successful teams are characterized by three factors: common ground, interpredictability, and directability \parencite{klein2005common}. Common ground includes the mutual knowledge, beliefs, and assumptions that enable interdependent action, encompassing both what is shared beforehand and what develops throughout the activity. As such, common ground is the prerequisite for interpredictability, which refers to the accurate anticipation of others’ actions. Skilled teams achieve this through shared knowledge and coordination practices developed over time. Finally, directability is the capacity to assess and adapt others’ actions as conditions change. Successful coordination depends on participants’ responsiveness and mutual commitment to addressing problems as they arise \parencite{bradshaw2017human, klein2005common}.

These factors can be applied to individual teammates to delineate the attributes of an effective agent in joint activities. \textcite{bradshaw2017human} identified eight characteristics that define such an agent: (1) observability---making one’s own current state and intentions transparent, (2) monitorability---enabling progress appraisal and signaling potential issues, (3) communicativeness---being informative, polite, and context-aware, (4) self-assessment---knowing one’s own limitations and balancing initiative with guidance, (5) reliability---being predictable and dependable, (6) directability---remaining responsive to external guidance and authority, (7) selectivity---focusing attention on what is most important, and (8) coordination---managing dependencies and maintaining common ground.

Following joint activity theory, an agent---whether human or artificial---that embodies these capabilities has the potential to contribute to the success of a joint activity. During interaction, an agent’s behavior serves as the key information source for perceiving it as a cooperative partner, since the relevant characteristics become directly observable in the activity and thus allow an assessment of an agent's cooperative potential. Yet potential alone is insufficient; it must be translated into actual cooperative behavior during the interaction itself. Both aspects are captured in the concept that we refer to as \emph{perceived cooperativity}: the perceived capability and cooperative practice of a partner to contribute to a successful joint activity, which becomes particularly relevant in synchronic human-AI integration.

\subsection{From evolutionary cooperation theory to teaming perception}
While perceived cooperativity captures the subjective perception of a cooperation partner's qualities within a single joint activity (reflecting the synchronic dimension of human–AI integration), interactions with AI systems increasingly extend beyond single episodes into repeated collaborations and long-term partnerships, making the diachronic dimension of integration equally relevant. Given that even machines without human-like features can be perceived as social actors \parencite{nass1994computers}, and that this tendency is amplified in AI systems with natural language capabilities \parencite{devrio2025linguistic, leeExploringDimensionsHuman2025}, the perception of AI as a teammate rather than a tool is an empirically grounded possibility rather than a mere assumption \parencite{attigMoreTaskPerformance2024, Lindgren16122024, walliserPerceptionTeamworkAutonomous2017}. Hence, at the diachronic level, the question is no longer only whether a cooperation partner is perceived as cooperative in a given moment, but whether the accumulated experience of repeated cooperative episodes gives rise to a sense of genuinely being a team \parencite{pacherieHowDoesIt2014}.

However, being part of a team entails more than the individual perception of another agent being a teammate. Teaming is a reciprocal relation that integrates a first-person experience of agency (i.e., being an active contributor whose actions matter for the outcome) with a global, collective framing of the activity as something \emph{we} are doing together \parencite{pacherieHowDoesIt2014}. Human–AI teaming thus relies on the simultaneous presence of first- and third-person perspectives, reflecting both (1) individual contributions of all actors and (2) shared, goal-directed actions. Together, the process of shared contribution of resources (e.g., time, cognitive effort, physical strength) towards mutually agreed goals is proposed to constitute the basis of the emergent state of teaming perception \parencite{marksTemporallyBasedFramework2001, riethUnveilingTeamEmergent2026}.

The emphasis on (1) individual contributions can be theoretically grounded in evolutionary cooperation theory, which defines cooperation through the costs and benefits that actions impose on individuals: Cooperative acts incur a cost to the actor while providing a benefit to others. The emergence and stability of such behavior depend on interaction structures that allow these individually costly acts to be favored by selection, for example through repeated interaction, relatedness, or reputation \parencite{nowak2006, westEvolutionaryExplanationsCooperation2007}. In this view, to cooperate is to incur a cost that enables another individual to obtain a benefit, stressing that a prosocial investment is the central defining characteristic of cooperative behavior \parencite{attigReciprocityPsychologicalNeeds2025, attig2026HCII}.


Shared, goal-directed action (2) is not merely the sum of individual efforts but a jointly constructed activity, making mutuality a structural feature of teaming. Genuinely shared action requires a collectively represented goal that both partners are jointly committed to pursuing \parencite{bratmanSharedCooperativeActivity1992a, tomasello2012twosteps}. This joint commitment requires each partner to actively support the other in performing their role and to deprioritize personal goals that might interfere with the shared objective \parencite{bratmanSharedCooperativeActivity1992a}. From this perspective, the human partner's own cooperative contributions, support, and willingness to set aside individual goals are constitutive of the teaming experience itself: A sense of genuine teamwork cannot emerge from the perception of AI contributions alone but requires the human to also experience themselves as an active, investing member of the team.

Importantly, for this experience of joint commitment and mutual support to arise, the non-human partner must itself be perceived as a goal-directed agent to a minimum degree. Without attributing some form of agency to the AI partner (i.e., without perceiving it as an entity that pursues goals, makes decisions, and actively contributes to the shared objective rather than merely producing outputs), the psychological conditions necessary for genuine teaming perception likely will not be met \parencite{musickWhatHappensWhen2021}. Perceived agency of the AI partner thus constitutes a prerequisite for teaming perception to emerge, and distinguishes cooperative AI agents from purely instrumental tools in the context of human-AI interaction \parencite{lyonsHumanAutonomyTeaming2021}.

We thus conceptualize teaming perception as a subjective state variable that emerges from the interplay of perceived AI prosocial contributions and mutual human–AI support. Based on this conceptualization, we organize the subjective measurement of \emph{teaming perception} into three facets: one capturing the emergent sense of the team as a whole, one capturing the human partner’s own cooperative investment and support, and one capturing the human’s perception of the AI partner’s supportiveness and contributions \parencite{attigMoreTaskPerformance2024}.

\subsection{Integrating perceived cooperativity and teaming perception}

Perceived cooperativity and teaming perception map onto complementary dimensions of the synchronic–diachronic integration framework (see Figure \ref{fig:illustration}): Perceived cooperativity is particularly sensitive to the synchronic dimension, assessing the quality of a single cooperative task sequence regardless of interaction history, while teaming perception extends into the diachronic dimension, capturing the relational experience that accumulates as repeated cooperative episodes build a joint history. Both constructs refer to the individual level and are process-oriented. That is, they target the perceptual and experiential indicators of successful cooperation rather than its outcomes.  
In this sense, the two scales together offer a measurement approach that tracks the subjective experience of human–AI cooperation across both its synchronic and diachronic dimensions.

The interplay between synchronic and diachronic human–AI integration also implies a developmental relationship between the two constructs: the accumulation of positive cooperative episodes (i.e., episodes characterized by high perceived cooperativity of the AI partner) likely consolidates expectations about the partner's ability to cooperate while also generating positive cooperative experience \parencite{attig2026HCII}, thereby giving rise to a genuine sense of teaming over time.

Notably, although teaming perception is theoretically most informative in the context of diachronic integration, it can also be meaningfully assessed after an initial encounter with an AI system in a cooperative setting, provided that the cooperative episode reaches a sufficient degree of synchronic integration. Under such conditions (i.e., high interdependency between partners to complete the joint task sequence), even a single cooperation instance needs contributions from both partners and enables the potential experience of mutual coordination and joint goal pursuit with the AI partner.

\section{Materials and methods}
\subsection{Scale construction}
The Perceived Cooperativity Scale (PCS) and the Teaming Perception Scale (TPS) were developed to assess two complementary aspects of human–AI cooperation: the perceived cooperativity of an AI partner within a given interaction, and the broader sense of teaming that emerges as cooperative episodes accumulate over time. To allow for comparisons between human–AI and human–human cooperation, adapted versions of both scales were additionally developed in which references to the AI agent (PCS-A, TPS-A) were replaced with references to a human partner (PCS-H, TPS-H).

Initial item pools were generated based on the theoretical constructs and prior research on cooperation and teaming. For the PCS, items were developed to capture the perceived cooperativity of an AI agent within a single cooperative task sequence. Based on the eight characteristics of a good agent for joint activity \parencite{bradshaw2017human}---observability, monitorability, communicativeness, self-assessment, reliability, directability, selectivity, and coordination---initial items were formulated to operationalize each characteristic from the perspective of the human user, targeting their subjective perception of the AI partner's behavior during a cooperative task sequence. Item formulations followed closely the descriptions by \textcite{bradshaw2017human}, while being deliberately rephrased in accessible, everyday language to ensure that participants without a technical background could respond to them intuitively. Moreover, we added the item stem \enquote{I had the feeling that} to account for the fact that many characteristics must be inferred rather than directly observed, framing all items consistently as first-person perception judgments rather than objective assessments.
For instance, the first characteristic reads as follows: \enquote{A good agent is observable. It makes its pertinent state and intentions obvious.} \parencite[p. 14]{bradshaw2017human}. This characteristic was transformed into two scale items: \enquote{I had the feeling that the agent could communicate its status to me.} and \enquote{I had the feeling that the agent made clear what it wanted to achieve.}. This approach was repeated for all characteristics. Items were iteratively refined in discussions among the authors to ensure content validity and item clarity. The pre-validation PCS comprised 15 items (see Tables \ref{tab:pcs-a} and \ref{tab:pcs-h} in Appendix \ref{app:pcs}). Item responses are provided on a 6-point Likert scale from 1 (\emph{completely disagree}) to 6 (\emph{completely agree}). 

For the TPS, item generation followed directly from the theoretical conceptualization of teaming perception as a subjective state variable emerging from the interplay of perceived AI prosocial contributions and mutual human–AI support \parencite{bratmanSharedCooperativeActivity1992a, pacherieHowDoesIt2014, tomasello2012twosteps}. The three facets underlying this conceptualization (i.e., the emergent sense of the team as a whole, one's own contribution and support, and the human's perception of the system's contribution and support) were each transferred into a set of 3 items (i.e., 9 items in total). As with the PCS, items were phrased in accessible, everyday language to ensure comprehensibility. Unlike the PCS, however, no additional item stem was needed, as TPS items are inherently formulated as first-person experiential judgments. Rather than targeting specific, potentially non-observable AI characteristics, they capture the respondent's holistic subjective experience of the interaction as a cooperative episode. Items were iteratively refined in discussions among the authors to ensure content validity and item clarity. The pre-validation TPS comprised 9 items (see Tables \ref{tab:tps-a} and \ref{tab:tps-h} in Appendix \ref{app:tps}). Item responses are provided on a 6-point Likert scale from 1 (\emph{completely disagree}) to 6 (\emph{completely agree}).

\subsection{Translations}
The translation of the German scales into English was conducted with the help of three native bilingual speakers. First, an initial translation was developed by the authors. Second, two bilingual speakers developed translations of the scales independently. Third, the first author discussed the translations with the other authors, resulting in one harmonized version of the scale translations with small adaptations. Fourth, the translated scales were given to the third bilingual speaker for blind translation back into German. Fifth, this back translation was checked against the original German versions by the first author. Seven instances were identified where the original scales and back translations across both scales were not in perfect accordance. Here, the English translations were adapted in consultation with the third bilingual speaker. 

\subsection{Scale evaluation procedure}
The resulting scales were administered to diverse samples of users participating in three independent studies on human-AI and human-automation collaboration. The results of participants' self reports were then used to assess item properties (item difficulty, discrimination and Cronbach's $\alpha$ if removed), scale dimensionality (exploratory and confirmatory factor analysis), reliability (Cronbach's $\alpha$ and McDonald's $\omega$), and construct validity (discriminant validity, ability to differentiate between systems). This approach is in line with the best practices of scale evaluation \parencite{boateng2018best}.

\subsubsection{Validity evaluation of the perceived cooperativity scale}

For discriminant validity of the PCS, we selected measures that tap conceptually adjacent but distinct aspects of human–AI interaction. While we expect positive correlations with all variables, differences in effect sizes are expected to reflect the varying degrees of conceptual overlap with perceived cooperativity.

Particularly strong correlations are expected for subjective information processing awareness (SIPA; \cite{schrills2023users}), which captures the degree to which a system is perceived as transparent, understandable, and predictable. Particularly items PCS01, PCS02, and PCS03 share conceptual overlap with the SIPA transparency subscale, while item PCS09 mirrors the SIPA subscale predictability. However, while SIPA focuses more narrowly on the cognitive, information-processing dimension of system perception, the PCS adopts a broader approach that additionally encompasses the agent's responsiveness to user input and its active management of the joint activity. Therefore, a perfect overlap between the two measures is not expected. Moreover, substantial correlations are expected for perceived competence \parencite{leach2007group} and system trustworthiness \parencite{franke2015advancing}, as both reflect core aspects of an agent's capability and reliability that are central to cooperative potential. 

By contrast, weaker positive correlations are expected for measures that are conceptually more distal from perceived cooperativity. System acceptance \parencite{van1997simple}, although potentially influenced by cooperative quality, reflects a broader evaluative attitude toward the system as a whole rather than the episodic perception of cooperative capability and practice, and is therefore expected to show a lower correlation with the PCS. Similarly, perceived usefulness \parencite{attig2025LLM} is included as a further indicator, capturing the broader instrumental dimension of the agent's actions. Finally, perceived warmth \parencite{leach2007group} captures the interpersonal and relational quality of the interaction---a dimension that is more closely tied to the experiential sense of teaming than to the assessment of cooperative capability that defines perceived cooperativity---and is therefore expected to show a weaker correlation with the PCS than the capability- and reliability-focused measures.

For evaluating the scale's ability to differentiate between cooperation partners with high and low perceived cooperativity, we compare scale mean scores for cooperation partners in a cooperative game setting that differ with respect to partner type (human vs. rule-based system vs. reinforcement learning based-system; see Study 1 and \cite{attigMoreTaskPerformance2024}). These three variations differ regarding human-likeness, transparency, and the degree to which they embody the characteristics of an effective cooperative agent as defined by joint activity theory as the basis for item formulation. Specifically, the human partner is expected to score highest on perceived cooperativity, as human partners naturally possess the full range of cooperative capabilities. The rule-based system is expected to score higher than the reinforcement learning-based system, because its behavioral rules were derived from human cooperative game strategies, making its actions more transparent, predictable, and comprehensible to the human partner. The reinforcement learning-based system, by contrast, developed cooperative game strategies autonomously through experience \parencite{canaanBehavioralEvaluationHanabi2020}, resulting in game styles that are not as easily interpretable by human partners, thereby limiting the human's ability to form accurate expectations and assess the agent's cooperative potential. If the PCS successfully differentiates between cooperation partners of varying cooperative quality, mean scores should reflect this ordering, with human partners rated highest, followed by the rule-based system, and the reinforcement learning-based system rated lowest.

\subsubsection{Validity evaluation of the teaming perception scale}

For discriminant validity of the TPS, we assessed perceived warmth and competence \parencite{leach2007group}. Perceived warmth captures the interpersonal and relational quality of the interaction and is expected to align with the prosocial and mutual support facets of teaming perception. Perceived competence is included because teaming perception requires that both partners are perceived as contributors whose efforts meaningfully advance the shared goal. An agent perceived as incapable is less likely to be experienced as a genuine teammate regardless of relational quality. 

By contrast, weaker positive correlations are expected for measures that are conceptually more distal from teaming perception. SIPA's narrower focus on cognitive transparency and predictability \parencite{schrills2023users} makes it conceptually distant from the relational, experiential quality of teaming perception: The cooperative engagement that teaming requires extends well beyond what information processing awareness can capture. Moreover, system usefulness reflects the broader instrumental dimension of the agent's actions \parencite{attig2025LLM}, capturing whether its outputs are practically valuable and facilitate task performance rather than the reciprocal and goal-directed experience of genuinely acting as a team, and is therefore also expected to show a weaker correlation with the TPS. Importantly, we expect stronger correlations with the partner-focused and team-focused facets of the TPS than with the self-focused facet, as the human partner's own cooperative investment and support reflect primarily their own motivation and goal prioritization rather than their perception of the AI's qualities.

For evaluating the scale's ability to differentiate between cooperation partners that vary in their potential for teaming perception, we compare scale mean scores across the same three partner types in a cooperative game setting (human vs. rule-based system vs. reinforcement learning-based system; see Study 1 and \cite{attigMoreTaskPerformance2024}). The expected ordering mirrors that of the PCS: The human partner is expected to score highest, as human partners are most naturally suited to engage in genuinely reciprocal cooperation and mutual support. The rule-based system is expected to score higher than the reinforcement learning-based system, as its human-derived behavioral strategies enable the human partner to develop more stable expectations about its game style and should therefore enable a smoother user experience and a sense of joint agency \parencite{pacherieHowDoesIt2014}. In contrast, the reinforcement learning-based system, whose less interpretable game style limits the formation of such expectations, is expected to score lowest. 

\subsection{Samples and study designs}
To enable analyses, we utilized data from three studies with various human-AI and human-automation applications, totaling $N$ = 409 participants. Details on the studies---labeled by the application examined---are reported below. Table \ref{tab:participants} shows key descriptive sample statistics. Beyond demonstrating the heterogeneity of use cases for the scales, the samples also differed greatly in age, gender, and affinity for technology interaction. Regarding the latter, we applied the Affinity for Technology Interaction (ATI) scale \parencite{frankePersonalResourceTechnology2019}, which evaluates a person’s self-reported tendency to engage actively in technology interaction and is rated on a response scale from 1 to 6. Samples are compared with a population distribution sample ($M$ = 3.61, $SD$ = 1.08, $N$ = 232; \cite{frankePersonalResourceTechnology2019}).   

\subsubsection{Study 1: Playing the cooperative card game Hanabi}
Study 1 used a within-subjects design to examine human-AI teaming using the cooperative card game \textit{Hanabi}, played in dyads of human participants or dyads consisting of a human and one of two artificial agents. Hanabi is a cooperative game in which players jointly build sequences of correctly ordered cards without seeing their own cards, relying solely on limited hints from their partners \parencite{attigMoreTaskPerformance2024}. Hanabi represents a particularly demanding challenge for AI because it inherently requires collaborative play and reasoning about a partner's mental state, managed through restricted communicative actions \parencite{bardHanabiChallengeNew2020}. This makes it not only a benchmark for AI development but also a prototypical setting for investigating human-AI teaming, as it fulfills the two requirements considered necessary to establish a joint activity \parencite{bradshaw2017human, klein2005common}: When entering the game, both players implicitly agree to collaborate towards a shared goal (i.e., scoring a maximum of points) through interdependent actions (i.e., game moves that are interlinked between players). These two requirements are inherent and non-negotiable characteristics of the game; thus, Hanabi can only be played cooperatively \parencite{bardHanabiChallengeNew2020}. Nevertheless, it allows for meaningful variation in perceived cooperativity and teaming perception, as partners can adopt different playing styles (e.g., by playing more indirectly by giving hints that reduce uncertainty, or taking a more active, direct role in achieving success by playing cards) creating the conditions necessary to observe and measure differences in cooperative quality across partner types.

Playing Hanabi with artificial agents provides coverage of both dimensions of human-AI integration: the highly interdependent nature of in-game actions represents a high degree of synchronic integration, while the repeated interactions across two games per partner introduce at least a limited diachronic dimension (compared to naturalistic long-term partnerships with varying tasks over time). Nevertheless, it allows for the beginning of accumulated cooperative experience to emerge within each partner condition.

Study 1 was conducted using a web-based interface for the game application \parencite{HRIHanabi}. Communication was restricted to predefined in-game hints to maintain standardization. Participants ($N$ = 94) played six rounds of Hanabi in total: two rounds each with another human player and with each of two artificial agents. The artificial agents were (1) the rule-based agent \enquote{Piers} (in the following referred to as RB, for rule-based), which was derived from human game play \parencite{walton-riversEvaluatingModellingHanabiPlaying2017}, and (2) the reinforcement learning agent \enquote{Rainbow} (in the following referred to as RL, for reinforcement learning), which learned to play Hanabi by playing with the RB agent \parencite{canaanBehavioralEvaluationHanabi2020}. RL agents that are trained with single game partners typically perform well when paired with their training partners, but this performance does not transfer to other game partners (i.e., fewer correct moves, more mistakes, higher rate of lost games; \cite{canaanBehavioralEvaluationHanabi2020}). Prior research has demonstrated that humans evaluate Hanabi agents that display dominated actions (i.e., moves that lead to loss of playable cards or life tokens) more negatively \parencite{siuPursuitPredictiveModels2025}. Taken together, we expect the RL agent to display less transparent and less predictable behavior than the RB agent, and therefore to elicit lower ratings of perceived cooperativity and teaming perception.

Data collection was split into two sessions: one session in which participants played two games with a human partner, and one session in which they played two games each with both artificial agents. The order of sessions and the order of artificial agents within the agent session were partially counterbalanced across participants, with the constraint that both artificial agent interactions were completed within the same session. This resulted in four possible orderings: (1) human-human first followed by the RB then RL agent, (2) human-human first followed by the RL then RB agent, (3) RB then RL agent followed by human-human, and (4) RL then RB agent followed by human-human. After each interaction (i.e., two rounds of Hanabi with one partner) participants completed the PCS and TPS alongside additional scales assessing construct validity. As Study 1 was the first study in which the current scale versions were tested, we additionally asked participants to rate their confidence in responding to the single PCS and TPS items on a scale from 1 (\emph{not confident at all}) to 10 (\emph{extremely confident}) to assess the item fit to the type of interaction. Partner type (human vs. artificial agent) was disclosed to participants after completing the two games but before the assessment of subjective variables, ensuring that knowledge of partner type would not influence participants' game experience. Conditions are referred to throughout this paper by abbreviated labels (i.e., Han-HH for human partner, Han-RB for rule-based artificial agent, Han-RL for the reinforcement learning agent).

Participants consisted of undergraduate and graduate students who were offered course credit and that were recruited via email, student forums, and personal invitations in lectures and seminars. The sample's ATI did not significantly differ from the general population distribution ($t$(93) = -2.55, $p$ = .012, $d$ = 0.26, small effect size). Ethical approval was obtained from the Ethics Committees of University of Lübeck (tracking number: 2023-570, July 17, 2023) and the Honda R\&D Bioethics Committee (tracking number: 100HM-015H, July 21, 2023). The study was conducted in accordance with the Declaration of Helsinki; all participants provided written informed consent prior to taking part. Study 1 builds on a pilot study reported in \textcite{attigMoreTaskPerformance2024}, extending it with a larger sample and the refined scale versions.

\subsubsection{Study 2: Large-language model interaction}
Study 2 used a within-subjects design to examine how AI autonomy and goal alignment influence user experience in the context of LLM use \parencite{attig2025LLM}. Compared to Hanabi in Study 1, this setting represents a lower degree of synchronic human-AI integration, as the fact-checking task could in principle be completed through a single exchange rather than requiring tightly interdependent actions, while repeated trials per chatbot version introduce a limited diachronic integration. 

Participants ($N$ = 182) fact-checked encyclopedic articles on natural and sociocultural phenomena using four versions of an LLM-based chatbot, implemented via GPT-4o Mini and iteratively refined to ensure alignment with experimental conditions. The chatbot versions differed systematically in their degree of autonomy and alignment with the user's goal of correcting factual errors: \textit{Focused} (supported the user goal without additional actions), \textit{Augmented} (supported the user goal and autonomously provided sources for corrections), \textit{Conflicted} (supported the user goal but also autonomously simplified the text, which conflicted with the user goal), and \textit{Failed} (did not support the user goal but implied capability of correcting spelling errors). This systematic variation in cooperative quality---from a fully aligned and capable partner to one that fails to contribute meaningfully to the shared goal---provides the conditions necessary to examine whether the PCS and TPS are sensitive to differences in perceived cooperativity across interaction contexts.

Participants engaged with each chatbot across twelve fact-checking trials (three trials per chatbot version), and the PCS, TPS, and additional construct validity measures were assessed after each condition. Participants were recruited via Prolific \parencite{prolific2024} and financially compensated according to Prolific's recommendations for ethical payment. The sample's ATI differed significantly from the general population ($t$(181) = 6.92, $p$ < .001, $d$ = 0.51, medium effect size). Ethical approval was obtained from the Ethics Committees of University of Lübeck (tracking number: 2024-452; August 28, 2024) and the Honda R\&D Bioethics Committee (tracking number: 101HM-028H; August 23, 2024). The study was conducted in accordance with the Declaration of Helsinki; all participants provided written informed consent prior to taking part. Conditions are referred to throughout this paper by their respective labels (e.g., Focused indicates data from the Focused chatbot condition). As the present work focuses on scale development and validation, a detailed report of the experimental results (including ability to differentiate between cooperation partners with high and low perceived cooperativity and teaming perception) is provided elsewhere \parencite{attig2025LLM}.

\subsubsection{Study 3: Long-distance charging and trip planner interaction}
Study 3 used a correlational design to examine perceived cooperativity in the context of real-world interaction with an algorithmic decision-support system, focusing on user interaction with electric vehicle (EV) trip and charging stop planning systems \parencite{stattkus2025LDC}. Study 3 targeted users of the Tesla trip planner and Tesla Supercharger network. This setting differs fundamentally from the prior two studies in the nature of the human-system interaction it involves. Rather than a cooperative AI agent that actively contributes to a shared goal through interdependent actions, the Tesla trip planner is best characterized as a decision-support system: It processes inputs and returns optimized routing and charging recommendations based on real-time data, but exercises no salient agency (\url{https://www.tesla.com/features/navigation}). The TPS was not assessed in Study 3 because the system's lack of agency means that the conditions necessary for teaming perception to emerge (i.e., mutual contribution, reciprocal support, and the prioritization of a shared goal by both partners) are not sufficiently met in this interaction context \parencite{lyonsHumanAutonomyTeaming2021}. This makes Study 3 a deliberate boundary case for the application of the PCS, allowing us to examine whether the scale remains meaningful in contexts where the interaction partner is an algorithmic tool rather than a cooperative agent.

Participants ($N$ = 133) were recruited at Tesla Superchargers in northern Germany as well as through an online forum on electric mobility. Participants were compensated with EUR 10,00 for their participation. The sample's ATI was significantly higher than the general population distribution ($t$(132) = 16.44, $p$ < .001, $d$ = 1.43, large effect size). Ethical approval was obtained from the ethics committee of the University of Lübeck (tracking number 2024-462; August 26, 2024); the Honda R\&D Bioethics Committee determined that ethical approval was not applicable for this study. The study was conducted in accordance with the Declaration of Helsinki; all participants provided written informed consent prior to taking part.

\begin{table}[h]
    \centering
    \caption{Sample demographics and application contexts}
    \renewcommand{\arraystretch}{1.2}
    \begin{tabular}{l c l l l c c c}
        \toprule
         Study & {\textit{N}} & Population & Application & Samples & Age \textit{M (SD)} & ATI \textit{M (SD)} & Gender (M:F:D) \\
        \midrule
        1: Hanabi & 94  & Students  & \makecell[l]{Cooperative\\card game}  & \makecell[l]{Han-HH\\Han-RB\\Han-RL}  & 23.0 (4.8)  & 3.4 (1.0)  & 19:74:1 \\
        \midrule
        2: LLM    & 182 & Prolific  & \makecell[l]{LLM-based\\chatbot}           & \makecell[l]{Augmented\\Focused\\Conflicted\\Failed} & 38.5 (13.2)  & 4.0 (0.8)  & 91:89:1 \\
        \midrule
        3: LDC    & 133 & EV Drivers  & \makecell[l]{EV trip\\planner}      & LDC  & 46.2 (10.8)  & 4.9 (0.9)  & 127:6:0 \\
        \bottomrule
    \end{tabular}
    \label{tab:participants}
\end{table}

\subsection{Construct validity measures}
If not otherwise stated, participants responded to measures on a 6-point Likert scale from 1 (\emph{completely disagree}) to 6 (\emph{completely agree}). As reliability indicators, we report Cronbach's $\alpha$ as a range across the conditions for each sample, which is interpreted according to common practice (e.g., \cite{cripps2017psychometric}) as poor ($0.5 \leq \alpha < 0.6$), questionable ($0.6 \leq \alpha < 0.7$), acceptable ($0.7 \leq \alpha < 0.8$), good ($0.8 \leq \alpha < 0.9$), or excellent ($\alpha \geq 0.9$).

\subsubsection{Subjective information processing awareness} To determine to what extent a system was perceived as transparent, understandable and predictable, the 9-item Subjective Information Processing Awareness Scale (SIPA, \cite{schrills2023users}) was used. Internal consistency was good to excellent ($\alpha$ = .85 - .92). It was assessed in Hanabi RB/RL and all LLM samples.  

\subsubsection{Perceived system trustworthiness} We assessed the perceived trustworthiness of the chatbot with the 5-item Facets of System Trustworthiness scale (FOST, \cite{franke2015advancing}). Internal consistency was acceptable to excellent ($\alpha$ = .79 - .90). It was assessed in the LLM and LDC samples. 

\subsubsection{User acceptance} The 9-item System Acceptance Scale \cite{van1997simple} assesses system acceptance along two dimensions: usefulness, which reflects the perceived effectiveness and practical benefits of the system, and affective satisfaction, which captures the user's emotional response to system interaction. Participants rated each item on a bipolar 5-point response scale, where each point represents opposing adjective pairs (e.g., "useless -- useful", "annoying -- pleasant"). Internal consistency was good ($\alpha$ = .80 - .81). It was assessed in the LDC sample. 

\subsubsection{System usefulness} For assessing usefulness \parencite{attig2025LLM}, we used four self-constructed items (\enquote{The outputs of the chatbot are not meaningful.}, \enquote{My decisions are facilitated when the system displays information.}, \enquote{I can use the chatbot’s output for my decisions.}, \enquote{I cannot assess what role the chatbot’s output plays in the task.}). Internal consistency was questionable to acceptable ($\alpha$ = .67 - .72). It was assessed in the LLM samples. 

\subsubsection{Perceived warmth and competence} \textcite{leach2007group} developed a scale to assess social perception in human-human interaction with 9 items on a 7-point response scale (from 1 -- \textit{strongly disagree} to 7 -- \textit{strongly agree}). We utilized the subscales perceived warmth and competence for assessing social perception of both human and artificial cooperation partners. Internal consistency was good to excellent ($\alpha$ = .80 - .97). Social perception was assessed in all Hanabi and all LLM samples.

\section{Results}
\subsection{Perceived cooperativity scale}
\subsubsection{Dimensionality}
Perceived cooperativity characterizes the perception of agents being equipped with requirements for cooperative teamwork according to joint activity theory \parencite{bradshaw2017human,klein2005common}. The items were developed to reflect a single underlying construct of perceived cooperativity. To examine the dimensionality of the PCS empirically, EFAs were conducted on the Hanabi samples of Study 1. First, parallel analysis using principal axis factoring (PAF) with 100 Monte Carlo simulations was employed to determine the number of factors. Results indicated a one-factor solution, further supported by the Kaiser criterion, which confirmed that only the first factor had an eigenvalue exceeding 1. To inspect item-level factor structure, a subsequent PAF with Promax rotation was conducted. Factor loadings were generally strong, ranging from .44 to .87. An exception was PCS14, which showed weaker loadings in the Han-RB (.28) and Han-HH (.32) conditions while performing adequately in the Han-RL condition (.60), suggesting some inconsistency in how this item functions across contexts. The single factor explained between 47~\% (Han-RL) and 55~\% (Han-HH) of the variance.

Confirmatory factor analyses (CFA) assessed model fit across all eight samples of Studies 1, 2, and 3 (Table \ref{tab:pcs_cfa}), comparing a 15-item one-factor model to a 14-item one-factor model (excluding PCS14, also based on item analysis results). Model fit was evaluated using Comparative Fit Index (CFI), Tucker–Lewis Index (TLI), Root Mean Square Error of Approximation (RMSEA), Standardized Root Mean Square Residual (SRMR), Akaike information criterion (AIC), and Bayesian information criterion (BIC). Following \textcite{browneAlternativeWaysAssessing1992}, RMSEA values above .10 were considered indicative of unacceptable fit. For CFI and TLI, values of .90 and above were treated as reflecting acceptable fit \parencite{bentlerSignificanceTestsGoodness1980}, and SRMR values below .08 were considered acceptable \parencite{huCutoffCriteriaFit1999}. Consistent with \textcite{marshSearchGoldenRules2004}, these thresholds were applied as approximate guidelines rather than fixed decision rules, with fit indices interpreted jointly and in light of the broader pattern of evidence. Moreover, recommended cut-offs are intended for the evaluation of a single model in isolation rather than the comparative fit of nested models \parencite{huCutoffCriteriaFit1999}.

Comparing the two models, the 14-item solution generally showed improved fit indices across most samples. Significant $\chi^2$ differences in Han-RL, LDC, Focused, Augmented, and Conflicted indicated that PCS14 negatively impacted model fit, while no meaningful difference was found for the remaining samples. Across both models, CFI met the .90 threshold in most samples, with the notable exception of the Augmented condition, where fit was poor across all indices. TLI values indicated poor fit of the 15-item model across most samples, and a rather inconclusive pattern for the 14-item model. RMSEA values were almost consistently within the acceptable range across models, while SRMR indices all were at least acceptable. This pattern was further supported by AIC and BIC values, which are information-theoretic indices that allow for direct model comparison irrespective of sample size, with lower values indicating better model fit relative to model complexity. AIC and BIC consistently favored the 14-item solution. Taken together, and interpreted as approximate guidelines rather than fixed decision rules \parencite{marshSearchGoldenRules2004}, the CFA results are consistent with a unidimensional factor structure across most contexts and an improved fit of the 14-item model. However, differences across samples suggest that the empirical factor structure of the PCS might be to some degree context- and interaction-dependent.

\begin{table}[htbp]
\centering
\footnotesize
\renewcommand{\arraystretch}{1.2}
\caption{Comparison of 14-item and 15-item CFA models}
\label{tab:pcs_cfa}
\setlength{\tabcolsep}{1.25pt}
\resizebox{\textwidth}{!}{%
\begin{tabular}{lccccccccccccc>{\centering\arraybackslash}p{0.8cm}}
\toprule
 & \multicolumn{6}{c}{14-item model} & \multicolumn{6}{c}{15-item model} & \multicolumn{2}{c}{Comparison} \\
\cmidrule(lr){2-7} \cmidrule(lr){8-13} \cmidrule(lr){14-15}
Sample & CFI & TLI & RMSEA & SRMR & AIC & BIC & CFI & TLI & RMSEA & SRMR & AIC & BIC & $\Delta\chi^2 (13)$ & $p$ \\
\midrule
Han-HH & .93 & .92 & 0.09 & 0.05 & 3516.02 & 3587.23 & .93 & .92 & 0.09 & 0.05 & 3827.42 & 3903.72 & 15.52 & .276 \\
Han-RB & .91 & .89\textsuperscript{2} & 0.10 & 0.06 & 3638.23 & 3709.44 & .90 & .88\textsuperscript{2} & 0.10 & 0.06 & 3915.54 & 3991.84 & 19.17 & .118 \\
Han-RL & .93 & .92 & 0.08 & 0.07 & 3534.73 & 3605.94 & .91 & .89\textsuperscript{2} & 0.09 & 0.07 & 3779.70 & 3856.00 & 29.27 & .006 \\
LDC       & .89\textsuperscript{1} & .87\textsuperscript{2} & 0.09 & 0.07 & 3605.94 & 5525.34 & .88\textsuperscript{1} & .86\textsuperscript{2} & 0.09 & 0.07 & 5844.91 & 5931.62 & 22.91 & .043 \\
Focused & .92 & .91 & 0.09 & 0.05 & 6753.30 & 6843.01 & .91 & .89\textsuperscript{2} & 0.10 & 0.05 & 7303.84 & 7399.96 & 32.44 & .002 \\
Augmented & .80\textsuperscript{1} & .77\textsuperscript{2} & 0.12\textsuperscript{3} & 0.08 & 6551.10 & 6640.82 & .77\textsuperscript{1} & .74\textsuperscript{2} & 0.13\textsuperscript{3} & 0.08 & 7127.25 & 7223.37 & 46.72 & <.001 \\
Conflicted & .91 & .90 & 0.10 & 0.05 & 6380.18 & 6469.90 & .90 & .88\textsuperscript{2} & 0.10 & 0.05 & 6922.49 & 7018.61 & 38.91 & <.001 \\
Failed  & .95 & .94 & 0.10 & 0.04 & 7170.03 & 7259.74 & .95 & .94 & 0.09 & 0.04 & 7625.39 & 7721.51 & 15.76 & .262 \\
\bottomrule
\multicolumn{15}{l}{\emph{Note.} \textsuperscript{1} CFI < .90, indicating poor fit. \textsuperscript{2} TLI < .90, indicating poor fit. \textsuperscript{3} RMSEA > .10, indicating unacceptable fit.}\\
\end{tabular}
}
\end{table}

\subsubsection{Item analysis}
\begin{figure}[htbp]
    \centering
    \includegraphics[width=\linewidth]{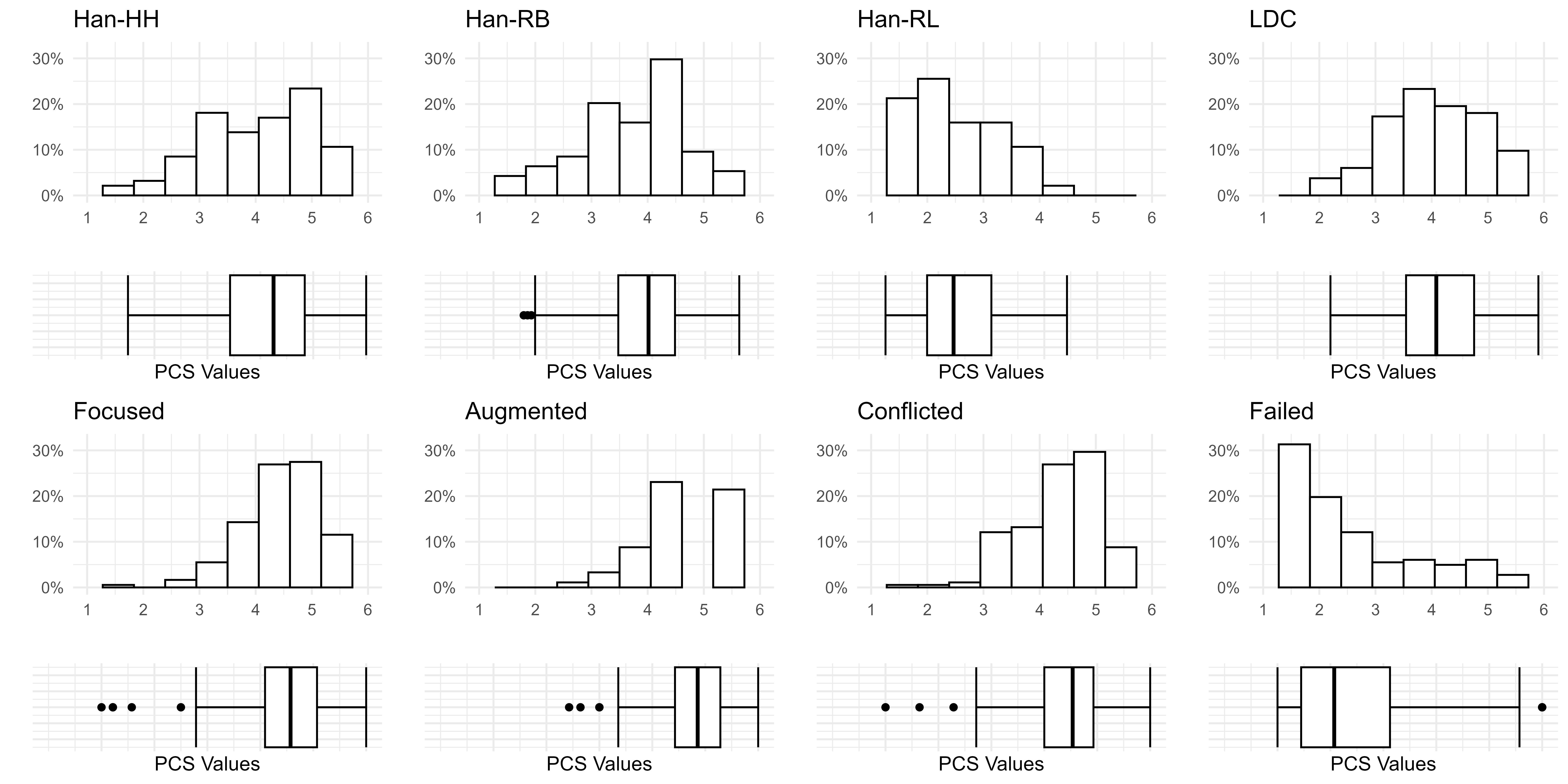}
    \caption{Histogram of PCS mean scores by sample. Boxplots below represent interquartile range (IQR), vertical lines are median, error bars are 1.5 times the IQR, and black dots are outliers beyond this threshold.}
    \label{fig:pcs_distrub}
\end{figure}

Item analysis (i.e., item discrimination, item difficulty, $\alpha$ if item removed, and confidence rating $z$-scores) was conducted on the Hanabi samples to examine scale performance on the single item level (Table \ref{tab:pcs_item_analysis}). Item discrimination coefficients (i.e., part-whole corrected item-total correlations) depict how well a single item differentiates between interaction ratings in the same direction as the overall scale. Item discrimination ranged on average across samples between $\bar{r}_{itc}$ = .39 (PCS14) to .81 (PCS15), indicating good item discrimination \parencite{moosbrugger2020}, with the exception of item PCS14 in the Han-HH ($r_{itc}$ = .32) and Han-RB samples ($r_{itc}$ = .28). Item difficulty indices reflect the average degree of agreement across the response scale, ranging from 0 to 100 \%, with higher values indicating that participants more readily agreed with the item on average. Item difficulty values ranged on average between 40.0 \% (PCS14) to 56.1 \% (PCS12), indicating adequate item difficulty \parencite{moosbrugger2020}. Results further revealed systematic differences between samples: The average difficulty value in the Han-RL condition (28.1 \%) was considerably lower than in the Han-HH (61.5 \%) and Han-RB (54.9 \%) conditions, suggesting that participants agreed the PCS items more readily when interacting with humans and the rule-based agent than with the reinforcement learning agent. This pattern becomes also apparent in the item scores (see Table \ref{tab:pcs_item_analysis} and Figure \ref{fig:pcs_distrub}) and will be further analyzed in Section 4.1.4. Cronbach's $\alpha$ if item removed showed that all items, except PSC14, contributed positively to scale reliability. 

Mean confidence ratings ranged from $M$ = 6.1 to $M$ = 7.8 (overall $M$ = 7.0), with standard deviations between $SD$ = 1.7 and $SD$ = 2.7 ($M_{\text{SD}}$ = 2.1). To identify items with relatively low confidence, item-level means and standard deviations were averaged across Hanabi samples and standardized using $z$-scores. Items were flagged if their mean $z$-score was more than 2 standard deviations below the overall mean or if their standard deviation $z$-score exceeded +1 standard deviation. This approach accounts for sample-specific response tendencies. Two items were flagged: PCS03 (high variability) and PCS14 (high variability and below-average confidence).

Taking CFA and item analysis results into account, we decided to exclude item PCS14 from the scale. Thus, all further analyses were performed on the 14-item version of the PCS.



\begin{landscape}
\begin{table}[H]
\centering
\caption{PCS item analysis metrics (based on Hanabi samples of Study 1)}
\label{tab:pcs_item_analysis}
\renewcommand{\arraystretch}{1}
\begin{tabular}{l ccc ccc ccc ccc cc}
\toprule
 & \multicolumn{3}{c}{Value $M$($SD$)} & \multicolumn{3}{c}{Item discrimination ($r_{itc}$)} & \multicolumn{3}{c}{Item difficulty} & \multicolumn{3}{c}{$\alpha$ if item removed} & \multicolumn{2}{c}{Confidence $z$} \\
\cmidrule(lr){2-4} \cmidrule(lr){5-7} \cmidrule(lr){8-10} \cmidrule(lr){11-13} \cmidrule(lr){14-15}
Item & Han-HH & Han-RB & Han-RL & Han-HH & Han-RB & Han-RL & Han-HH & Han-RB & Han-RL & Han-HH & Han-RB & Han-RL & $M$ & $SD$ \\
\midrule
PCS01 & 3.6 (1.3) & 3.4 (1.2) & 2.4 (1.0) & .73 & .62 & .65 & 52.7 & 48.9 & 27.5 & .94 & .93 & .92 & -1.0 & 0.4 \\
PCS02 & 3.8 (1.2) & 3.7 (1.2) & 2.3 (1.1) & .80 & .73 & .61 & 56.6 & 53.6 & 25.5 & .94 & .93 & .92 & 0.0 & -0.3 \\
PCS03 & 3.6 (1.2) & 3.3 (1.2) & 2.1 (1.1) & .80 & .72 & .68 & 52.8 & 46.4 & 22.6 & .94 & .93 & .92 & -0.8 & 1.1\textsuperscript{5} \\
PCS04 & 4.2 (1.2) & 3.4 (1.2) & 2.2 (1.1) & .74 & .70 & .78 & 64.0 & 48.3 & 24.3 & .94 & .93 & .92 & -0.1 & -0.5 \\
PCS05 & 4.2 (1.3) & 3.8 (1.3) & 2.5 (1.3) & .78 & .66 & .65 & 64.5 & 55.5 & 30.6 & .94 & .93 & .92 & 0.0 & 0.3 \\
PCS06 & 3.9 (1.3) & 3.7 (1.3) & 2.3 (1.1) & .73 & .73 & .72 & 58.7 & 54.5 & 26.4 & .94 & .93 & .92 & -0.7 & 0.1 \\
PCS07 & 4.4 (1.2) & 3.9 (1.2) & 2.9 (1.2) & .73 & .52 & .62 & 68.5 & 57.9 & 37.2 & .94 & .93 & .92 & -0.2 & 0.1 \\
PCS08 & 4.4 (1.3) & 4.0 (1.2) & 2.7 (1.2) & .73 & .72 & .71 & 67.7 & 60.2 & 33.4 & .94 & .93 & .92 & 0.2 & 0.2 \\
PCS09 & 3.7 (1.3) & 3.7 (1.3) & 2.3 (1.3) & .64 & .71 & .45 & 53.6 & 53.6 & 26.6 & .94 & .93 & .93 & 0.5 & -1.1 \\
PCS10 & 4.4 (1.2) & 4.0 (1.4) & 2.1 (1.1) & .75 & .75 & .71 & 67.5 & 59.8 & 21.7 & .94 & .93 & .92 & 0.8 & -0.1 \\
PCS11 & 4.4 (1.2) & 3.9 (1.2) & 2.8 (1.3) & .57 & .54 & .53 & 67.0 & 58.9 & 36.8 & .94 & .93 & .92 & 0.3 & -0.6 \\
PCS12 & 4.6 (1.3) & 4.3 (1.2) & 2.6 (1.3) & .82 & .70 & .61 & 71.7 & 65.5 & 31.1 & .94 & .93 & .92 & 2.1 & -2.0 \\
PCS13 & 4.2 (1.3) & 3.9 (1.3) & 2.2 (1.2) & .71 & .76 & .70 & 63.2 & 58.9 & 24.0 & .94 & .93 & .92 & 0.9 & -0.3 \\
PCS14 & 3.4 (1.3) & 3.1 (1.1) & 2.5 (1.1) & .32\textsuperscript{1} & .28\textsuperscript{1} & .56 & 48.1 & 41.9 & 30.0 & .95\textsuperscript{2} & .94\textsuperscript{2} & .92 & -2.4 \textsuperscript{4} & 2.6\textsuperscript{5} \\
PCS15 & 4.3 (1.4) & 3.9 (1.4) & 2.2 (1.1) & .80 & .83 & .80 & 66.4 & 58.7 & 24.5 & .94 & .92 & .92 & 0.4 & 0.0 \\
\midrule
Scale & 4.1 (1.0) & 3.7 (0.9) & 2.4 (0.8) & .71 & .66 & .65 & 61.5 & 54.9 & 28.1 & .94\textsuperscript{3} & .93\textsuperscript{3} & .92\textsuperscript{3} &  &  \\
\bottomrule
\end{tabular}
\par\smallskip
\begin{minipage}{\linewidth}
\footnotesize\emph{Note.} \textsuperscript{1} Item discrimination $<$ .4. \textsuperscript{2} $\alpha$ if item removed $>$ overall $\alpha$. \textsuperscript{3} Scale Cronbach's $\alpha$. \textsuperscript{4} Below-average z-standardized mean confidence ratings ($z$ $<$ $-2$ $SD$). \textsuperscript{5} High variability of z-standardized $SD$ of confidence ratings ($z$ $>$ 1 $SD$).
\end{minipage}
\end{table}
\end{landscape}

\subsubsection{Reliability}
Reliability analysis of the PCS was conducted on all samples (see Table \ref{tab:pcs_tps_reliability}). The PCS yielded on average internal consistency values of Cronbach's $\alpha$ = .93 and McDonalds $\omega$ = .93. According to common guidelines for interpretation (e.g., \cite{cripps2017psychometric}), these values indicate excellent internal consistency, suggesting that the PCS is a highly reliable measure across diverse contexts of human-AI cooperation.

\subsubsection{Validity}
Table \ref{tab:perception_correlations} shows the correlations of the PCS with construct validity indicators. All effect sizes were interpreted according to \cite{cohen1992power}. We conducted multiple Shapiro-Wilk tests to assess normality and found that most variables were non-normally distributed. Therefore, we only report Spearman correlations for subsequent analyses.

All correlations between the PCS and construct validity indicators were positive, as expected. Effect sizes followed the anticipated pattern: The strongest correlations emerged for perceived competence ($\bar{r}_S$ = .74), SIPA ($\bar{r}_S$ = .68), and system trustworthiness ($\bar{r}_S$ = .66). Indicators expected to correlate more weakly---system acceptance (${r}_S$ = .64), perceived warmth ($\bar{r}_S$ = .57), and perceived usefulness ($\bar{r}_S$ = .53)---nonetheless showed strong associations.


\begin{table}[htbp]
\centering
\renewcommand{\arraystretch}{1.2}
\caption{Reliability for PCS and TPS (composite and subscales) across samples}
\label{tab:pcs_tps_reliability}
\begin{tabular}{lcccccccccc}
\toprule
 & \multicolumn{2}{c}{PCS} & \multicolumn{8}{c}{TPS} \\
\cmidrule(lr){2-3} \cmidrule(lr){4-11}
 & & & \multicolumn{2}{c}{\makecell{Composite\\[-3pt]scale}} & \multicolumn{2}{c}{Team} & \multicolumn{2}{c}{Self} & \multicolumn{2}{c}{Partner} \\
\cmidrule(lr){2-3} \cmidrule(lr){4-5} \cmidrule(lr){6-7} \cmidrule(lr){8-9} \cmidrule(lr){10-11}
Sample & $\alpha$ & $\omega$ & $\alpha$ & $\omega$ & $\alpha$ & $\omega$ & $\alpha$ & $\omega$ & $\alpha$ & $\omega$ \\
\midrule
Han-HH     & .95 & .95 & .86 & .88 & .89 & .90 & .40 & .61 & .69 & .73 \\
Han-RB     & .94 & .94 & .84 & .85 & .88 & .88 & .50 & .57 & .78 & .79 \\
Han-RL     & .92 & .93 & .81 & .84 & .90 & .91 & .37 & .42 & .80 & .81 \\
Focused    & .94 & .94 & .92 & .92 & .92 & .92 & .67 & .73 & .79 & .83 \\
Augmented  & .90 & .90 & .88 & .90 & .91 & .91 & .57 & .67 & .63 & .74 \\
Conflicted & .94 & .94 & .90 & .91 & .92 & .92 & .58 & .73 & .78 & .81 \\
Failed     & .96 & .96 & .93 & .94 & .96 & .96 & .78 & .79 & .95 & .95 \\
LDC        & .91 & .91 &    &    &    &    &    &    &    &    \\
\bottomrule
\end{tabular}
\end{table}

\begin{table}[H]
\centering
\caption{Spearman correlations (and \textit{p}-values) between PCS subscales and validity indicators across samples}
\label{tab:perception_correlations}
\renewcommand{\arraystretch}{1.5}
\begin{tabular}{lcccccccc}
\toprule
Variable & Han-HH & Han-RB & Han-RL & LDC & Augmented & Focused & Conflicted & Failed \\
\midrule
SIPA       & & \makecell{.76\\[-3pt](<.001)} & \makecell{.62\\[-3pt](<.001)} & & \makecell{.65\\[-3pt](<.001)} & \makecell{.69\\[-3pt](<.001)} & \makecell{.62\\[-3pt](<.001)} & \makecell{.74\\[-3pt](<.001)} \\
Warmth     & \makecell{.55\\[-3pt](<.001)} & \makecell{.59\\[-3pt](<.001)} & \makecell{.43\\[-3pt](<.001)} & & \makecell{.50\\[-3pt](<.001)} & \makecell{.64\\[-3pt](<.001)} & \makecell{.62\\[-3pt](<.001)} & \makecell{.68\\[-3pt](<.001)} \\
Competence & \makecell{.84\\[-3pt](<.001)} & \makecell{.76\\[-3pt](<.001)} & \makecell{.67\\[-3pt](<.001)} & & \makecell{.65\\[-3pt](<.001)} & \makecell{.70\\[-3pt](<.001)} & \makecell{.76\\[-3pt](<.001)} & \makecell{.78\\[-3pt](<.001)} \\
Acceptance & & & & \makecell{.64\\[-3pt](<.001)} & & & & \\
FOST       & & & & \makecell{.57\\[-3pt](<.001)} & \makecell{.58\\[-3pt](<.001)} & \makecell{.68\\[-3pt](<.001)} & \makecell{.69\\[-3pt](<.001)} & \makecell{.76\\[-3pt](<.001)} \\
Usefulness & & & & & \makecell{.48\\[-3pt](<.001)} & \makecell{.47\\[-3pt](<.001)} & \makecell{.51\\[-3pt](<.001)} & \makecell{.66\\[-3pt](<.001)} \\
\bottomrule
\end{tabular}
\end{table}


To evaluate the PCS's ability to differentiate between cooperation partners of varying cooperative capabilities, we repeated the statistical approach of \textcite{attig2025LLM}---where differences in PCS mean scores between LLM-sampled Augmented, Focused, and Conflicted conditions had been tested---using the Hanabi samples from Study 1. Distributions of PCS scores across all samples are depicted in Figure \ref{fig:pcs_distrub}, including histograms and boxplots. A repeated-measures ANOVA comparing PCS mean scores across three partner types was conducted: a human partner (Han-HH), a rule-based system (Han-RB), and a reinforcement learning-based system (Han-RL). Mauchly's test indicated that the assumption of sphericity was met ($p$ = .194), and Shapiro-Wilk's test indicated that the residuals deviated from a normal distribution ($p$ = .049). However, visual inspection of the QQ-plot supported an approximately normal distribution. No outliers were detected (Grubbs test, $p$ = 1.00).

The ANOVA revealed a significant large main effect of partner type on PCS scores, $F(2, 186) = 92.26$, $p < .001$, $\eta_G^2 = .40$, indicating that perceived cooperativity differed substantially across conditions. A planned contrast analysis with weights ($-0.66$, $-0.33$, $1$) confirmed the hypothesized ordering Han-RL $<$ Han-RB $<$ Han-HH, $t(93) = 9.93$, $p < .001$, $d = 1.02$ (large effect). Pairwise comparisons with Bonferroni-Holm correction showed that PCS scores were significantly lower for Han-RL than for Han-RB, $t(93) = -11.39$, $p < .001$, $d = -1.22$, and lower for Han-RL than for Han-HH, $t(93) = -12.55$, $p < .001$, $d = -1.23$ (both large effects). The difference between Han-RB and Han-HH was also significant, $t(93) = -2.34$, $p = .021$, $d = -0.23$ (small effect). Taken together, these results support the hypothesized ordering and indicate that the PCS successfully differentiates between cooperation partners of varying cooperative capability.

\subsection{Teaming perception scale}
\subsubsection{Dimensionality}
Teaming perception characterizes the emergent state of cooperation partners functioning as a team, arising from the interplay of perceived AI prosocial contributions and mutual human–AI support. To assess its factor structure, the TPS was developed with a theoretically grounded 3-factor model, which we first examined with EFA to assess scale dimensionality before confirming it with CFA.

As with the PCS, EFAs were conducted on the Hanabi samples of Study 1. First, parallel analysis using principal axis factoring (PAF) with 100 Monte Carlo simulations was employed to determine the number of factors. Results were inconclusive, supporting a two-factor solution for Han-HH sample (explaining 60 \% of variance) and a three-factor solution for both Han-RB and Han-RL samples (explaining 65 \% and 60 \% of variance). However, based on the Kaiser criterion, all three Hanabi samples indicated a one-factor solution with only the first factor having an eigenvalue exceeding 1 (explaining between 44 \% and 49\% of variance).

To inspect item-level factor structure, subsequent PAFs with Promax rotation were conducted. Factor loadings for the one-, two- and three-factor solutions showed that the two-factor model did not yield an easily interpretable factor structure (Factor 1: TPS01, TPS02, TPS03, TPS04, TPS07, TPS09; Factor 2: TPS05, TPS06, TPS08). The empirical three-model solution differed from the theoretical one, but was interpretable: On average across samples, Factor 1 comprised Items TPS01, TPS02, TPS03, TPS07, and TPS09 (items referring to general perception of teaming and partner contributions), Factor 2 comprised Items TPS04 and TPS06 (items referring to one's own support of the partner and team), and Factor 3 comprised items TPS05 and TPS08 (items referring to shared goal prioritization). 


With CFAs, we assessed model fit of the one-factor solution, the three-factor solution based on EFA, and the three-factor solution based on theoretical considerations across the seven samples of Studies 1 and 2 (Table \ref{tab:tps_cfa}). Results indicate that the three-factor models showed consistently improved fit over a one-factor model across all samples, as indicated by significant $\chi^2$-difference tests. The relative fit of the two three-factor solutions, however, varied across samples: The EFA-based model showed superior fit in the Hanabi samples, which is unsurprising given that the EFA was conducted on these data. More notably, the theory-based model performed comparably or better across the LLM-based samples (Focused, Augmented, Conflicted, Failed), supporting the generalizability of the theoretical factor structure across different human–AI cooperation contexts. This pattern was further supported by AIC and BIC values, which are information-theoretic indices that allow for direct model comparison irrespective of sample size, with lower values indicating better model fit relative to model complexity. AIC and BIC consistently favored the EFA-based model in the Hanabi samples and the theory-based model in the LLM-based samples. Crucially, however, the differences in fit between the two three-factor solutions were nuanced rather than substantial.\footnote{Please note that in this case, the model comparison is limited to comparing fit indices focusing on AIC and BIC information criteria, as traditional likelihood ratio tests are not suitable due to the non-nested model structure of the two three-factor solutions \parencite{merkleTestingSEM2016}.} We therefore decided to retain the theory-based factor structure, as it is more directly grounded in our theoretical framework and yields the best interpretable subscales.

\begin{landscape}
\begin{table}[H]
\centering
\caption{Comparison of one-factor and three-factor CFA models}
\label{tab:tps_cfa}
\renewcommand{\arraystretch}{1.2}
\small
\setlength{\tabcolsep}{3pt}
\resizebox{\linewidth}{!}{%
\begin{tabular}{@{}lcccccccccccccccccccc@{}}
\toprule
 & \multicolumn{6}{c}{Three-factor model (theory)} & \multicolumn{6}{c}{Three-factor model (EFA)} & \multicolumn{4}{c}{One-factor model} & \multicolumn{2}{c}{Comparison (theory)} & \multicolumn{2}{c}{Comparison (EFA)} \\
\cmidrule(lr){2-7} \cmidrule(lr){8-13}  \cmidrule(lr){14-17} \cmidrule(lr){18-19} \cmidrule(lr){20-21}
Sample & CFI & TLI & RMSEA & SRMR & AIC & BIC & CFI & TLI & RMSEA & SRMR & AIC & BIC  & CFI & TLI & RMSEA & SRMR & $\Delta\chi^2$ (3) & $p$ & $\Delta\chi^2$ (3) & $p$ \\
\midrule
Han-HH & .91 & .87\textsuperscript{2} & 0.14\textsuperscript{3} & 0.08 & 2188.63 & 2242.04 & .97 & .95 & 0.08 & 0.04 & 2163.39 & 2216.80 & .86\textsuperscript{1} & .82\textsuperscript{2} & 0.16\textsuperscript{3} & 0.10\textsuperscript{4} & 43.79 & <.001 & 87.94 & <.001 \\
Han-RB & .94 & .92 & 0.10 & 0.08 & 2163.44 & 2216.85 & .99 & .99 & 0.04 & 0.06 & 2143.20 & 2196.61 & .90 & .87\textsuperscript{2} & 0.13\textsuperscript{3} & 0.09\textsuperscript{4} & 15.08 & .002 & 30.26 & <.001 \\
Han-RL & .95 & .93 & 0.10 & 0.07 & 2264.31 & 2317.72 & .94 & .92 & 0.10 & 0.06 & 2267.25 & 2320.66 & .91 & .88\textsuperscript{2} & 0.12\textsuperscript{3} & 0.08 & 8.43 & .038 & 10.44 & .015 \\
Focused & .97 & .95 & 0.09 & 0.06 & 4135.26 & 4202.54 & .92 & .88\textsuperscript{2} & 0.15\textsuperscript{3} & 0.05 & 4189.36 & 4256.64 & .90 & .82\textsuperscript{2} & 0.16\textsuperscript{3} & 0.06 & 38.60 & <.001 & 22.28 & <.001 \\
Augmented & .95 & .92 & 0.11\textsuperscript{3} & 0.09\textsuperscript{4} & 4131.40 & 4198.69 & .92 & .88\textsuperscript{2} & 0.13\textsuperscript{3} & 0.06 & 4157.04 & 4224.33 & .91 & .88\textsuperscript{2} & 0.13\textsuperscript{3} & 0.07 & 17.83 & <.001 & 5.27 & .153 \\
Conflicted & .93 & .90 & 0.13\textsuperscript{3} & 0.07 & 4025.71 & 4093.00 & .91 & .86\textsuperscript{2} & 0.15\textsuperscript{3} & 0.05 & 4051.33 & 4118.62 & .89\textsuperscript{1} & .85\textsuperscript{2} & 0.16\textsuperscript{3} & 0.06 & 35.41 & <.001 & 14.32 & .003 \\
Failed & .98 & .97 & 0.10 & 0.05 & 4342.13 & 4409.42 & .96 & .94 & 0.13\textsuperscript{3} & 0.06 & 4373.53 & 4440.81 & .92 & .89\textsuperscript{2} & 0.18\textsuperscript{3} & 0.10\textsuperscript{4} & 41.63 & <.001 & 48.08 & <.001 \\
\bottomrule
\end{tabular}
}
\par\smallskip
\begin{minipage}{\linewidth}
\footnotesize\emph{Note.} \textsuperscript{1} CFI $<$ .90, indicating poor fit. \textsuperscript{2} TLI $<$ .90, indicating poor fit. \textsuperscript{3} RMSEA $>$ .10, indicating unacceptable fit. \textsuperscript{4} SRMR $>$ .08, suggesting inadequate fit.
\end{minipage}
\end{table}
\end{landscape}

\subsubsection{Item analysis}

Item analysis followed the same procedure as for the PCS, with the exception that results are reported separately for the three TPS subscales (see Table \ref{tab:tps_item_analysis}).

For the Team subscale, item discrimination coefficients ranged on average across samples between $\bar{r}_{itc}$ = .75 (TPS03) and .79 (TPS01), indicating good item discrimination \parencite{moosbrugger2020}. Item difficulty values ranged on average between 55.8 \% (TPS01) and 61.6 \% (TPS03), reflecting adequate difficulty overall. As observed for the PCS, difficulty values were considerably lower in the Han-RL condition (35.2 \%) than in the Han-HH (73.3 \%) and Han-RB (67.5 \%) conditions, suggesting that participants agreed with Team subscale items more readily when interacting with the human partners and rule-based agent. Cronbach's $\alpha$ if item removed indicated that TPS02 and TPS03 would marginally increase subscale reliability in the Han-RL sample if removed. No items were flagged based on confidence rating $z$-scores.

For the Self subscale, item discrimination was substantially lower than for the Team and Partner subscales. Average item discrimination values across samples ranged between $\bar{r}_{itc}$ = .09 (TPS05) and .31 (TPS04), suggesting that the Self subscale measures a different construct than the overall scale. Item difficulty values ranged between 56.6 \% (TPS05), 71.7 \% (TPS04) and 80.6 \% (TPS06), indicating that participants tend to agree readily with TPS04 and TPS06 across samples. Cronbach's $\alpha$ if item removed revealed that TPS05 would marginally increase subscale reliability across all three samples if removed. No items were flagged based on confidence rating $z$-scores.

For the Partner subscale, item discrimination coefficients ranged on average across samples between $\bar{r}_{itc}$ = .48 (TPS08) and .75 (TPS07), indicating good discrimination overall \parencite{moosbrugger2020}. Item difficulty values ranged on average between 39.7 \% (TPS08) and 59.5 \% (TPS07), indicating adequate difficulty. As with the Team subscale, difficulty was considerably lower in the Han-RL condition (28.3 \%) than in Han-HH (67.5 \%) and Han-RB (57.3 \%). Cronbach's $\alpha$ if item removed indicated that TPS08 would marginally increase subscale reliability across all three samples if removed. TPS08 was flagged on both confidence indicators: It showed below-average mean confidence ($z$ = $-$2.26, more than 2\,$SD$ below the overall mean) and high variability in confidence ratings ($z$ = 2.28, more than 1\,$SD$ above the overall mean).

\begin{landscape}
\begin{table}[H]
\centering
\caption{TPS item analysis metrics (based on Hanabi samples of Study 1)}
\label{tab:tps_item_analysis}
\footnotesize
\renewcommand{\arraystretch}{1.2}
\begin{tabular}{l ccc ccc ccc ccc cc}
\toprule
 & \multicolumn{3}{c}{Value $M (SD)$} & \multicolumn{3}{c}{Item discrimination ($r_{itc}$)} & \multicolumn{3}{c}{Item difficulty} & \multicolumn{3}{c}{$\alpha$ if item removed} & \multicolumn{2}{c}{Confidence $z$} \\
\cmidrule(lr){2-4} \cmidrule(lr){5-7} \cmidrule(lr){8-10} \cmidrule(lr){11-13} \cmidrule(lr){14-15}
Item & Han-HH & Han-RB & Han-RL & Han-HH & Han-RB & Han-RL & Han-HH & Han-RB & Han-RL & Han-HH & Han-RB & Han-RL & $M$ & $SD$ \\
\midrule
TPS01 & 4.5 (1.5) & 4.3 (1.3) & 2.6 (1.3) & .79 & .84 & .81 & 69.2 & 66.0 & 32.1 & .83 & .79 & .75 & 0.44 & -0.43 \\
TPS02 & 4.9 (1.1) & 4.5 (1.3) & 2.9 (1.3) & .76 & .77 & .71 & 77.7 & 70.0 & 37.2 & .83 & .80 & .77\textsuperscript{2} & 0.27 & 0.07 \\
TPS03 & 4.7 (1.2) & 4.3 (1.2) & 2.8 (1.2) & .79 & .80 & .74 & 73.2 & 66.6 & 36.2 & .83 & .80 & .77\textsuperscript{2} & 0.13 & -0.53 \\
\midrule
Team Subscale & 4.7 (1.1) & 4.4 (1.1) & 2.8 (1.2) & .78 & .80 & .75 & 73.3 & 67.5 & 35.2 &  &  &  &  &  \\
\midrule
TPS04 & 4.8 (1.0) & 5.0 (0.8) & 4.8 (0.9) & .53 & .28\textsuperscript{1} & .13\textsuperscript{1} & 76.6 & 79.8 & 76.2 & .85 & .85 & .83 & 0.63 & -1.01 \\
TPS05 & 3.7 (1.3) & 3.9 (1.1) & 3.9 (1.3) & .15\textsuperscript{1} & .04\textsuperscript{1} & .07\textsuperscript{1} & 53.8 & 57.2 & 58.7 & .89\textsuperscript{2} & .87\textsuperscript{2} & .85\textsuperscript{2} & -0.92 & 0.64 \\
TPS06 & 5.0 (0.8) & 5.1 (0.7) & 5.0 (0.9) & .58 & .14\textsuperscript{1} & .19\textsuperscript{1} & 79.6 & 82.1 & 80.2 & .85 & .86 & .83 & 1.06 & -0.91 \\
\midrule
Self Subscale & 4.5 (0.7) & 4.7 (0.6) & 4.6 (0.7) & .42 & .15 & .13 & 70.0 & 73.1 & 71.7 &  &  &  &  &  \\
\midrule
TPS07 & 4.9 (1.2) & 4.4 (1.2) & 2.6 (1.2) & .75 & .79 & .71 & 77.5 & 68.3 & 32.8 & .83 & .80 & .77 & 0.40 & -0.23 \\
TPS08 & 3.7 (1.2) & 3.1 (1.1) & 2.2 (1.1) & .42 & .45 & .58 & 53.2 & 42.6 & 23.4 & .86\textsuperscript{2} & .83\textsuperscript{2} & .79\textsuperscript{2} & -2.26\textsuperscript{4} & 2.28\textsuperscript{5} \\
TPS09 & 4.6 (1.1) & 4.1 (1.3) & 2.4 (1.2) & .69 & .80 & .70 & 71.9 & 61.1 & 28.7 & .84 & .79 & .77 & 0.22 & 0.11 \\
\midrule
Partner Subscale & 4.4 (0.9) & 3.9 (1.0) & 2.4 (1.0) & .62 & .68 & .66 & 67.5 & 57.3 & 28.3 &  &  &  &  &  \\
\bottomrule
Composite Scale & 4.5 (0.8) & 4.3 (0.7) & 3.3 (0.7) & .61 & .55 & .52 & 70.3 & 66.0 & 45.1 & .86\textsuperscript{3} & .84\textsuperscript{3} & .81\textsuperscript{3} &  &  \\
\bottomrule
\end{tabular}
\par\smallskip
\begin{minipage}{\linewidth}
\footnotesize\emph{Note.} \textsuperscript{1} Item discrimination $<$ .4. \textsuperscript{2} $\alpha$ if item removed $>$ subscale $\alpha$. \textsuperscript{3} Scale Cronbach's $\alpha$. \textsuperscript{4} Below-average z-standardized mean confidence ratings ($z$ $<$ $-2$ $SD$). \textsuperscript{5} High variability of z-standardized $SD$ of confidence ratings ($z$ $>$ 1 $SD$).
\end{minipage}
\end{table}
\end{landscape}

\begin{figure}[H]
    \centering
    \includegraphics[width=\linewidth]{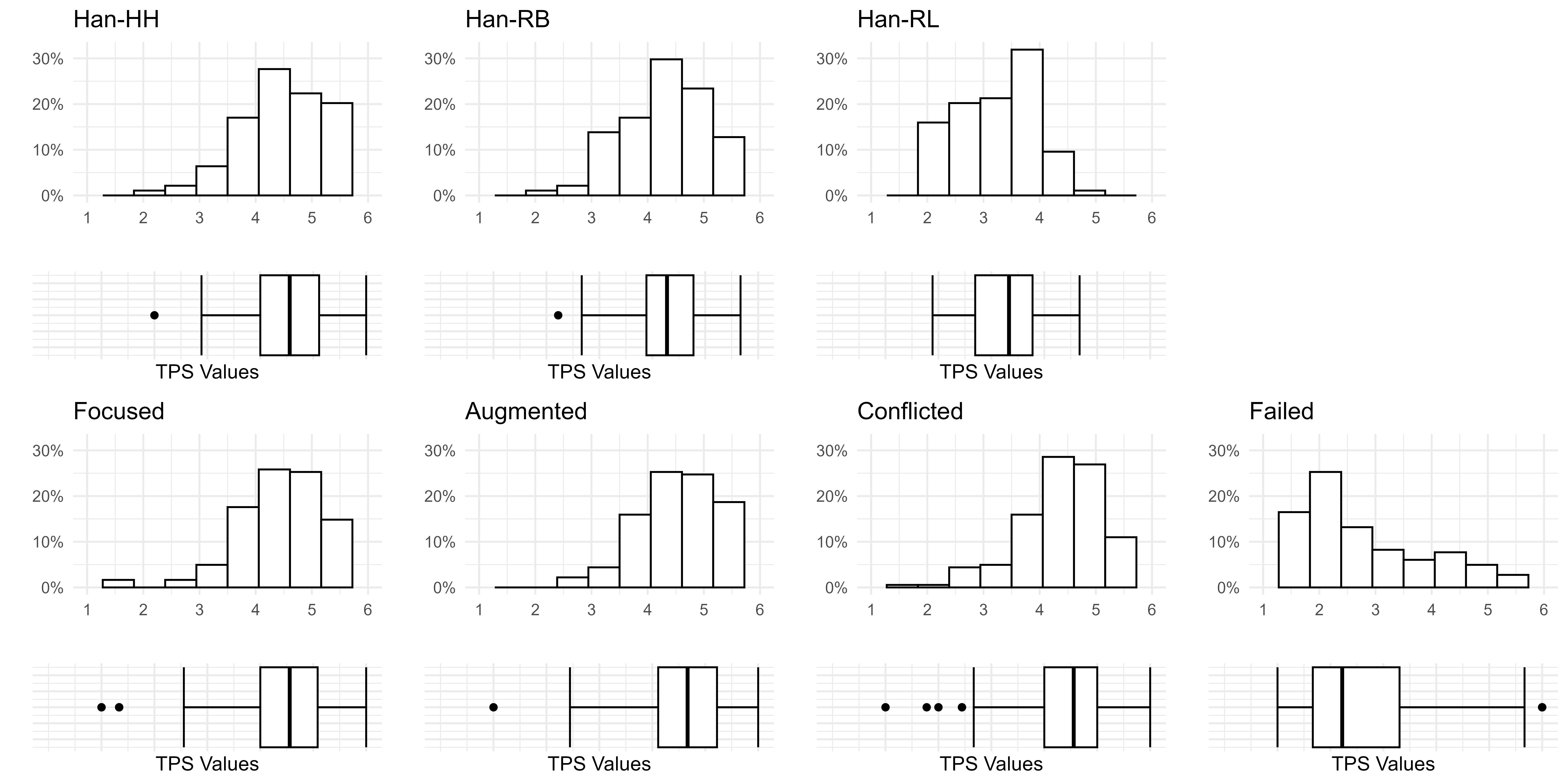}
    \caption{Histogram of TPS composite mean scores by sample. Boxplots below represent interquartile-range (IQR), vertical lines are median, error bars are 1.5 times the IQR, and black dots are outliers beyond this threshold.}
    \label{fig:TPS_distrub}
\end{figure}

\subsubsection{Reliability}
Reliability analysis of the TPS was conducted for samples of Studies 1 and 2 (see Table \ref{tab:pcs_tps_reliability}). All reliability coefficients reported below reflect averages across samples; interpretation followed common conventions \parencite{cripps2017psychometric}. The composite scale showed good internal consistency ($\alpha$ = .88, $\omega$ = .89). The Team subscale yielded excellent reliability ($\alpha$ = .91, $\omega$ = .91). The Partner subscale demonstrated acceptable to good reliability ($\alpha$ = .77, $\omega$ = .81). The Self subscale, in contrast, showed poor to questionable reliability ($\alpha$ = .55, $\omega$ = .65), with notably low values in the Hanabi samples. These results suggest that while the Team and Partner subscales are psychometrically robust across applications, the Self subscale shows higher internal consistency in non-game-based contexts.

\subsubsection{Validity}
For construct validation of the TPS, we used only Spearman correlations for subsequent analyses due to non-normality (as for the PCS). Table \ref{tab:tps_subscales_compact} shows the Spearman correlations between the TPS subscales and construct validity indicators across samples. All effect sizes were interpreted according to \textcite{cohen1992power}. 

The Team and Partner subscales showed a consistent pattern of strong positive correlations across all samples and indicators, broadly in line with expectations. As anticipated, the strongest associations emerged for perceived competence (Team: $\bar{r}_S$ = .66; Partner: $\bar{r}_S$ = .68). Perceived warmth also correlated strongly with both subscales (Team: $\bar{r}_S$ = .56; Partner: $\bar{r}_S$ = .57), albeit with slightly lower correlation coefficients. The conceptually more distal indicators---SIPA and system usefulness---showed, however, similarly sized effects (SIPA~--~Team: $\bar{r}_S$ = .56; Partner: $\bar{r}_S$ = .59; system usefulness~--~Team: $\bar{r}_S$ = .51; Partner: $\bar{r}_S$ = .55).

The Self subscale showed a markedly different and more heterogeneous pattern, with substantially weaker correlations across samples, consistent with theoretical expectations. Excluding Han-RL, mean correlations were moderate for PCS ($\bar{r}_S$ = .44), SIPA ($\bar{r}_S$ = .37), warmth ($\bar{r}_S$ = .37), and competence ($\bar{r}_S$ = .37), with system usefulness showing the weakest association ($\bar{r}_S$ = .26). In Han-RB, several individual associations were non-significant. In the Han-RL condition, negative correlations were found, specifically with perceived competence (${r}_S$ = -.36) and warmth (${r}_S$ = -.25).

To evaluate the PCS's ability to differentiate between cooperation partners of varying cooperative capabilities, we repeated the statistical approach of \textcite{attig2025LLM}---where differences in TPS composite mean scores between LLM-sampled Augmented, Focused, and Conflicted conditions had been tested---using the Hanabi samples from Study 1. Distributions of TPS composite scores across samples (Study 1, Study 2) are depicted in Figure \ref{fig:TPS_distrub}, including histograms and boxplots. A repeated-measures ANOVA comparing TPS composite mean scores across three partner types was conducted: a human partner (Han-HH), a rule-based system (Han-RB), and a reinforcement learning-based system (Han-RL). Mauchly's test indicated that the assumption of sphericity was met ($p$ = .718). Shapiro-Wilk's test indicated a violation of the normality assumption ($p$ = .018); however, visual inspection of the QQ-plot suggested approximate normality, and the ANOVA was conducted as planned given its robustness to mild normality violations. No outliers were detected (Grubbs test, $p$ = 1.00).

The ANOVA revealed a significant main effect of partner type on TPS scores, $F(2, 186) = 79.16$, $p < .001$, $\eta_G^2 = .36$, indicating that teaming perceptions differed substantially across conditions. A planned contrast analysis with weights ($-0.66$, $-0.33$, $1$) confirmed the hypothesized ordering Han-RL $<$ Han-RB $<$ Han-HH, $t(93) = 9.66$, $p < .001$, $d = 1.00$ (large effect). Pairwise comparisons with Bonferroni-Holm correction showed that TPS scores were significantly lower for Han-RL than for Han-RB, $t(93) = -9.90$, $p < .001$, $d = -1.05$, and lower for Han-RL than for Han-HH, $t(93) = -12.09$, $p < .001$, $d = -1.20$ (both large effects). The difference between Han-RB and Han-HH did not reach significance, $t(93) = -1.94$, $p = .056$, $d = -0.19$ (negligible effect). Taken together, these results largely support the hypothesized ordering and indicate that the TPS successfully differentiates between cooperation partners of varying cooperative capability. Distribution of TPS composite scores across samples (Study 1, Study 2) are depicted in Figure \ref{fig:TPS_distrub}.

\begin{table}[H]
\centering
\caption{Spearman correlations (and \textit{p}-values) between TPS subscales and validity indicators across samples}
\label{tab:tps_subscales_compact}
\renewcommand{\arraystretch}{1.5}
\resizebox{\textwidth}{!}{%
\begin{tabular}{l ccccccc}
\toprule
\multirow{2}{*}{Variable} & \multicolumn{7}{c}{Team subscale} \\
\cmidrule(lr){2-8}
& Han-HH & Han-RB & Han-RL & Focused & Augmented & Conflicted & Failed\\
\midrule
PCS 
& \makecell{.82\\[-3pt](<.001)} & \makecell{.83\\[-3pt](<.001)} & \makecell{.76\\[-3pt](<.001)} & \makecell{.68\\[-3pt](<.001)} & \makecell{.63\\[-3pt](<.001)} & \makecell{.69\\[-3pt](<.001)} & \makecell{.73\\[-3pt](<.001)}\\
SIPA 
& & \makecell{.66\\[-3pt](<.001)} & \makecell{.52\\[-3pt](<.001)} & \makecell{.63\\[-3pt](<.001)} & \makecell{.49\\[-3pt](<.001)} & \makecell{.45\\[-3pt](<.001)} & \makecell{.62\\[-3pt](<.001)}\\
Warmth 
& \makecell{.49\\[-3pt](<.001)} & \makecell{.59\\[-3pt](<.001)} & \makecell{.45\\[-3pt](<.001)} & \makecell{.57\\[-3pt](<.001)} & \makecell{.57\\[-3pt](<.001)} & \makecell{.57\\[-3pt](<.001)} & \makecell{.68\\[-3pt](<.001)} \\
Competence
& \makecell{.76\\[-3pt](<.001)} & \makecell{.73\\[-3pt](<.001)} & \makecell{.53\\[-3pt](<.001)} & \makecell{.62\\[-3pt](<.001)} & \makecell{.55\\[-3pt](<.001)} & \makecell{.67\\[-3pt](<.001)} & \makecell{.77\\[-3pt](<.001)}\\
System Usefulness
& &  &  & \makecell{.48\\[-3pt](<.001)} & \makecell{.39\\[-3pt](<.001)} & \makecell{.50\\[-3pt](<.001)} & \makecell{.66\\[-3pt](<.001)}\\
\midrule
\multirow{2}{*}{} & \multicolumn{7}{c}{Self subscale} \\
\cmidrule(lr){2-8}
PCS
& \makecell{.47\\[-3pt](<.001)} & \makecell{.16\\[-3pt](.130)} & \makecell{-.20\\[-3pt](.054)} & \makecell{.56\\[-3pt](<.001)} & \makecell{.58\\[-3pt](<.001)} & \makecell{.60\\[-3pt](<.001)} & \makecell{.42\\[-3pt](<.001)}\\
SIPA
& & \makecell{.16\\[-3pt](.113)} & \makecell{.05\\[-3pt](.660)} & \makecell{.52\\[-3pt](<.001)} & \makecell{.42\\[-3pt](<.001)} & \makecell{.40\\[-3pt](<.001)} & \makecell{.35\\[-3pt](<.001)}\\
Warmth
& \makecell{.25\\[-3pt](.017)} & \makecell{.07\\[-3pt](.482)} & \makecell{-.25\\[-3pt](.014)} & \makecell{.49\\[-3pt](<.001)} & \makecell{.53\\[-3pt](<.001)} & \makecell{.47\\[-3pt](<.001)} & \makecell{.39\\[-3pt](<.001)}\\
Competence
& \makecell{.33\\[-3pt](.001)} & \makecell{.09\\[-3pt](.392)} & \makecell{-.36\\[-3pt](<.001)} & \makecell{.43\\[-3pt](<.001)} & \makecell{.46\\[-3pt](<.001)} & \makecell{.50\\[-3pt](<.001)} & \makecell{.38\\[-3pt](<.001)}\\
System Usefulness
& &  &  & \makecell{.26\\[-3pt](<.001)} & \makecell{.25\\[-3pt](.001)} & \makecell{.26\\[-3pt](<.001)} & \makecell{.27\\[-3pt](<.001)}\\
\midrule
\multirow{2}{*}{} & \multicolumn{7}{c}{Partner subscale} \\ 
\cmidrule(lr){2-8}
PCS
& \makecell{.76\\[-3pt](<.001)} & \makecell{.77\\[-3pt](<.001)} & \makecell{.79\\[-3pt](<.001)} & \makecell{.77\\[-3pt](<.001)} & \makecell{.67\\[-3pt](<.001)} & \makecell{.71\\[-3pt](<.001)} & \makecell{.77\\[-3pt](<.001)} \\
SIPA
& & \makecell{.63\\[-3pt](<.001)} & \makecell{.55\\[-3pt](<.001)} & \makecell{.64\\[-3pt](<.001)} & \makecell{.56\\[-3pt](<.001)} & \makecell{.55\\[-3pt](<.001)} & \makecell{.61\\[-3pt](<.001)} \\
Warmth
& \makecell{.53\\[-3pt](<.001)} & \makecell{.64\\[-3pt](<.001)} & \makecell{.48\\[-3pt](<.001)} & \makecell{.58\\[-3pt](<.001)} & \makecell{.53\\[-3pt](<.001)} & \makecell{.53\\[-3pt](<.001)} & \makecell{.71\\[-3pt](<.001)} \\
Competence
& \makecell{.70\\[-3pt](<.001)} & \makecell{.70\\[-3pt](<.001)} & \makecell{.59\\[-3pt](<.001)} & \makecell{.71\\[-3pt](<.001)} & \makecell{.61\\[-3pt](<.001)} & \makecell{.68\\[-3pt](<.001)} & \makecell{.80\\[-3pt](<.001)} \\
System Usefulness
& &  &  & \makecell{.50\\[-3pt](<.001)} & \makecell{.43\\[-3pt](.001)} & \makecell{.55\\[-3pt](<.001)} & \makecell{.71\\[-3pt](<.001)}\\
\bottomrule
\end{tabular}
}
\end{table}


\section{Discussion}
\subsection{Summary of results}
The objective of the present research was to develop and validate two new scales to assess perceived cooperativity and teaming perception in human–AI cooperation, grounded in joint activity theory and evolutionary cooperation theory respectively. Tests of the scales across three studies with diverse samples and interaction contexts ($N = 409$) showed satisfying results with regard to dimensionality, reliability, validity, and the ability to differentiate between cooperation partners. Specifically, the results can be summarized as follows:

\begin{itemize}
\item Factor analyses indicated unidimensionality for the PCS and a three-factor structure for the TPS, with subscales reflecting team perception, self-contribution, and partner contribution.
\item Item analyses revealed good item discrimination and adequate item difficulty for the majority of items in both scales, with the Self subscale of the TPS showing notably weaker discrimination compared to the Team and Partner subscales.
\item Reliability analyses showed excellent internal consistency for the PCS across all samples and contexts, and good to excellent internal consistency for the TPS Team and Partner subscales, while the Self subscale showed acceptable internal consistency primarily in non-game-based contexts.
\item Construct validity analyses support expected relationships to subjective information processing awareness, perceived competence and warmth, system trustworthiness, system acceptance, and system usefulness, with effect sizes generally following the anticipated pattern of conceptual proximity to the respective constructs. Additionally, both scales successfully differentiated between cooperation partners of varying cooperative quality, with human partners rated highest, followed by the rule-based agent, and the reinforcement learning-based agent rated lowest, confirming the hypothesized ordering.
\item Boundary case analyses using the Tesla trip planner in Study 3 suggest that the PCS remains applicable in contexts of lower agent agency.
\end{itemize}

\subsection{Critical discussion of scale evaluation results}
While the overall psychometric properties of the PCS and TPS are promising, several limitations emerged from the analyses that warrant closer examination. For the PCS, the most striking psychometric issue was the inconsistent performance of item PCS14 across contexts. This item (\enquote{I had the feeling that the person/agent was able to realize that we had different approaches to achieving the goal.}) showed weaker item discrimination and lower confidence ratings than the remaining items. CFA results further suggested the 14-item solution without PCS14 to be  more appropriate than the 15-item solution. Consequently, we excluded the item from the scale, reducing the PCS to 14 items.

The correlational validity analyses of both the PCS and the TPS revealed that, while correlations with conceptually proximal constructs were strongest as hypothesized, some constructs that were expected to show weaker associations nonetheless yielded strong correlations. As \textcite{moradbakhtiDevelopmentValidationBasic2024} argue for a similar pattern of findings, these unexpectedly high correlations do not necessarily reduce the validity of the scales as such: A degree of correlation between psychological constructs is to be expected in the social sciences, as virtually all theoretically meaningful variables share some common variance \parencite{meehlWhySummariesResearch1990}. Moreover, given that all measures in the present studies were administered within the same session and in close succession, common method bias \parencite{podsakoffCommonMethodBiases2003} may have inflated some of the observed associations. Future studies should therefore further examine the discriminant validity of the PCS and TPS by using designs that reduce the risk of common method bias (e.g., by separating the administration of the focal scales and the validity indicators across time points). Additionally, criterion validity should be established by linking scale scores to objective indicators of cooperative success (e.g., task performance, error rates) as well as to implicit measures of cooperative experience (e.g., spontaneous use of first-person plural language when describing the interaction, the degree to which participants follow or override AI suggestions, indices of behavioral synchrony between human and AI partner actions over time).

Regarding the TPS, the most striking psychometric issues concern the Self subscale. Unlike the Team and Partner subscales, which ask participants to evaluate the interaction and the AI partner from a third-person perspective, the Self subscale turns the evaluative lens inward, asking participants to assess their own cooperative contribution and support from a first-person perspective. This fundamental difference in both content and perspective means that the Self subscale is, by design, measuring something distinct from the other two subscales. The substantially lower item discrimination coefficients observed across the three Hanabi conditions in Study 1 are consistent with this: Low discrimination values indicate that the Self subscale measures a different aspect of the construct than the other two subscales \parencite{moosbrugger2020}, which is theoretically expected. However, when also taking the item difficulty values into account, a more revealing pattern emerges. Items TPS04 and TPS06 showed consistently high difficulty values across conditions, suggesting that participants tended to rate themselves as highly cooperative, regardless of the interaction partner they were paired with. This gives rise to the hypothesis that the Self subscale---although designed as an interaction-specific self evaluation measure---may, under certain conditions, capture a more stable and positively biased evaluation of one's own cooperative tendency rather than a genuine reflection of the specific interaction (possibly prone to social desirability bias; \cite{chungExploringSocialDesirability2003}). Particularly telling in this regard are the raw item score means (see Table \ref{tab:tps_item_analysis}): While the Team and Partner subscale scores decreased systematically across conditions in line with the decreasing cooperative quality of the interaction partner, the Self subscale mean scores remained stable across all three conditions (Han-HH: 4.5, Han-RB: 4.7, Han-RL: 4.6), providing further support in the raw data that participants' self-evaluations were largely unaffected by the nature of the actual interaction.

Supporting this interpretation, negative correlations between the Self subscale and agent evaluations such as perceived warmth and competence emerged in the Han-RL condition: When the interaction partner had lower perceived cooperativity, participants still rated their own cooperative investment highly, producing an inverse relationship between self- and partner-focused evaluations. At the same time, the structural characteristics of the Hanabi task likely contributed to this pattern. Because the game enforces cooperative behavior structurally (i.e., players cannot opt out of cooperating with their game partner), there is comparatively restricted variance regarding to what extent participants can meaningfully vary their cooperative investment across conditions. The Self subscale items may therefore have captured a genuine psychological reality (i.e., participants did feel they were contributing and supporting their partner), but without sufficient variability to produce strong psychometric indices. Taken together, both interpretations imply one conclusion: When the task structure restricts the degrees of freedom regarding one's own cooperative investment, what remains visible is a tendency to evaluate oneself positively regardless of the specific interaction. Consistent with this interpretation, internal consistency of the Self subscale improved in the LLM-based interaction context of Study 2, where participants had considerably more freedom in deciding how actively to engage with the AI partner, and where larger variation in cooperative investment was therefore possible. These findings suggest that the Self subscale, in its current form, may be sensitive to interaction context when sufficient behavioral freedom exists, but susceptible to stable positivity bias when structural constraints limit that freedom. 

Another psychometric issue of the TPS concerns the two items related to goal prioritization (TPS05 and TPS08). Item analysis revealed near-zero discrimination values for TPS05 across all three Hanabi conditions, and notably low confidence ratings and high response variability for TPS08. Moreover, both items loaded onto a shared factor in the EFAs for the Han-RB and Han-RL samples, suggesting that, under certain conditions, they form a distinct dimension that is not well integrated with the remaining TPS items in these contexts. Specifically, genuinely prioritizing a shared goal over individual goals might not have been a meaningful experience in the Hanabi setting, which is a context where game partners share an identical goal by design, leaving little room for individual goals to compete with shared ones. As a consequence, participants' responses to TPS05 may have reflected a self-serving evaluation of their own cooperative tendency rather than a genuine interaction-specific experience, while TPS08 might have confronted participants with a judgment that was simply not answerable given the constraints of the task. Together, these findings point to a context-dependency in the performance of the goal prioritization items that should be taken into account when applying the TPS in future research.

Further, these findings suggest that the Self subscale in its current form likely requires refinement to more reliably capture interaction-specific cooperative experience rather than stable self-evaluations. At the same time, the observed pattern may itself carry theoretical value, as it raises the question of how sensitive individuals' self-perceived cooperative investment is to their actual cooperation experience, and if this sensitivity may play an important role in shaping how people attribute responsibility for cooperation failures, adapt their behavior to poorly performing AI partners, or calibrate their trust in AI systems over time.

Finally, it is worth noting that all three validation studies relied on relatively short interaction episodes, limiting the degree to which diachronic integration could be realized. While the Hanabi study introduced a limited diachronic element through repeated game rounds, and the LLM study included multiple trials per chatbot version, none of the studies examined genuine long-term partnerships characterized by deep diachronic integration over a variety of tasks. Future studies might therefore adopt longitudinal designs to examine how perceived cooperativity and teaming perception develop, stabilize, and interact over time, whether early cooperative experiences predict later teaming perception, and whether the Self subscale proves more sensitive and reliable when participants have had sufficient opportunity to develop a genuine sense of their own role and investment in a long-term human-AI partnership.

\subsection{Recommendations for scale use}

The following recommendations are intended to guide researchers in applying the PCS and TPS in a way that is informed by both the theoretical framework and the psychometric findings of the present validation studies.

\begin{itemize}

\item \textbf{Assess the interaction context and choose the appropriate scale(s).} Before selecting a scale, assess the degree of synchronic and diachronic integration of the target interaction. The PCS is appropriate across a wide range of contexts, including first encounters, relatively low synchronic integration, and systems with limited agency. When the primary interest concerns the perceived cooperative quality of the AI agent and study resources are limited, the PCS alone is recommended: Given its psychometric robustness and unidimensional factor structure, it can be easily administered and scored, making it a economical option for applied evaluation purposes. The TPS is appropriate when a minimum degree of synchronic integration and agent agency are present and the shared task is genuinely cooperative (i.e., when the task requires meaningful contributions from both partners and a sense of joint goal pursuit can plausibly emerge). When the research question calls for a more complete picture of the cooperative experience (e.g., when examining how single cooperative sequences and long-term teaming interact), both scales should be administered.

\item \textbf{Conduct a pilot study.} Where feasible, pre-test the scales with a small sample prior to main data collection, including item-level confidence ratings (e.g., by asking participants to rate their confidence in responding to single items on a scale from 1 [\emph{not confident at all}] to 10 [\emph{extremely confident}], as we did in Study 1). Pay particular attention to items whose applicability may depend on context-specific features: For instance, TPS05 and TPS08 (goal prioritization) may not be meaningful in settings where individual and shared goals cannot genuinely diverge, and PCS15 might not be applicable when no problems arose during the interaction). Items flagged by low confidence ratings or high response variability should be considered for omission before the main study.

\item \textbf{Administer the scales at the right time.} In repeated-measures or longitudinal designs, administer the scales after each interaction episode rather than retrospectively---particularly when the cooperation partners differ between episodes (e.g., LLM-based chatbots with different system prompts; \cite{attig2025LLM}). Cooperative experience is likely episode-specific, and retrospective ratings risk conflating distinct episodes into a single undifferentiated judgment.

\item \textbf{Examine the factor structure.} For the TPS, conduct exploratory and, where sample size permits, confirmatory factor analyses on the collected data. The theory-based three-factor structure (Team, Self, Partner) should be used as the default, particularly when the empirical factor structure is inconclusive or when the theoretically proposed subfacets show improved model fit. Retaining the theory-based structure ensures cross-study comparability. However, if the data clearly supports another factor structure, calculate subscale scores for the empirical factors.

\item \textbf{Report and interpret subscales appropriately.} Report the three TPS subscales separately, as they capture meaningfully distinct aspects of the teaming experience. A composite TPS score may be reported as a broad indicator of overall teaming perception, but we would recommend to only do so if the internal consistency of the composite reaches at least a good level (Cronbach's $\alpha \geq .80$). The Self subscale should be interpreted with particular caution given its susceptibility to positivity bias and context-dependent psychometric properties.

\end{itemize}


\subsection{Implications and conclusion}
The present research provides evidence to conclude that the PCS and TPS are promising tools for assessing the subjective experience of human–AI cooperation on both synchronic and diachronic dimensions. Together, the scales address a gap in the existing measurement landscape: While a large variety of scales has been developed to assess general attitudes towards AI \parencite{schepmanMeasurementAttitudesArtificial2025} and AI literacy \parencite{lintnerSystematicReviewAI2024}, theory-based instruments capable of capturing the episodic, process-oriented experience of cooperative interaction with AI agents are more rare. By grounding the PCS in joint activity theory \parencite{bradshaw2017human, klein2005common} and the TPS in evolutionary cooperation theory \parencite{nowak2006, tomasello2012twosteps} as well as shared action frameworks \parencite{bratmanSharedCooperativeActivity1992a, pacherieHowDoesIt2014}, the scales allow for connecting empirical findings to established literature on cooperation and joint activity. In the following, we will highlight specific implications from the present results.

\paragraph{Informing AI system design through diagnostic items.} Rather than capturing users' overall disposition towards AI or their general satisfaction with a system, the proposed scales target the perceived quality of a specific cooperative episode and the emergent sense of teaming that arises from it. This specificity makes them sensitive to differences that broader instruments may miss, as evidenced by the scales' ability to discriminate between cooperation partners that differ in their cooperative quality. Viewed from an applied perspective, the PCS in particular has diagnostic potential beyond its use as an outcome measure. Because its items were derived directly from the eight characteristics of an effective cooperative agent as defined by joint activity theory (i.e., observability, monitorability, communicativeness, self-assessment, reliability, directability, selectivity, and coordination; \cite{bradshaw2017human}), subscale or item-level scores can reveal which specific cooperative capabilities an AI agent is perceived to lack, enabling actionable design feedback. For instance, if users consistently rate an agent low on items related to predictability and directability, but higher on items related to communicativeness, this pattern points to specific aspects of agent behavior that designers should prioritize for enhanced user experience and acceptance. The TPS complements this by capturing whether the overall interaction gives rise to a genuine sense of teaming, including the human partner's own sense of contribution and mutual support. Thus, including the TPS in system evaluations might be particularly beneficial in highly interdependent human-AI cooperation settings where both human and artificial contributions determine task success. For instance, in industrial human-robot collaboration, where the quality of the shared outcome depends on both partners continuously adapting to each other's actions, a positive overall task outcome may mask an unsatisfying cooperative experience \parencite{siuEvaluationHumanAITeams2021}, and the TPS provides a means to capture this experiential dimension independently of performance metrics.

\paragraph{Enabling cross-agent comparisons.} A practical benefit of the scales is the availability of both AI and human partner versions, which use identical item formulations with only the referent changed. This allows to directly benchmark human-AI cooperation against human-human cooperation in a given context using the same instrument, enabling more nuanced conclusions about whether and to what degree AI agents can serve as genuine cooperation partners. Such comparisons are particularly relevant in hybrid team settings where humans collaborate with both human and artificial teammates.


\paragraph{Applicability to diverse study designs.} The PCS and TPS are sufficiently economical to be embedded in larger study designs without imposing undue participant burden. They can serve as dependent variables for evaluating AI system quality, as manipulation checks in experimental designs that vary cooperative conditions, or as moderator and mediator variables in models of human-AI interaction outcomes. The availability of both AI and human partner versions further extends their utility to studies with mixed or hybrid team designs. Given the heterogeneity of the three samples in the present validation studies (i.e., ranging from students playing a cooperative card game to Prolific participants interacting with LLM-based chatbots to electric vehicle drivers using a real-world trip planning system), the scales appear suited for deployment across a wide range of populations and application contexts.

\section*{Disclosure of potential conflicts of interest}
All authors declare that they have no conflicts of interest.

\section*{Author contributions: CRediT statement}
Christiane Attig: Conceptualization; Funding acquisition; Investigation; Methodology; Project administration; Resources; Visualization; Writing – original draft; Writing – review \& editing. Christiane Wiebel-Herboth: Conceptualization; Project administration; Resources; Supervision; Writing – review \& editing. Patricia Wollstadt: Conceptualization; Resources; Writing – review \& editing. Tim Schrills: Conceptualization; Methodology. Mourad Zoubir: Formal Analysis; Visualization; Writing – original draft. Thomas Franke: Conceptualization; Funding acquisition; Project administration; Resources; Supervision; Writing – review \& editing. 						

\section*{Acknowledgments}
We want to thank Laura Duwe, Sophia von Elbwart, Simon Flintrop, Bennet Hut, and Jacob Mellin for their support in collecting data for the Hanabi study as well as Beate Stattkus-Fortange for the LDC study. Further, we want to thank Tobias Harms, Jan Heidinger, Maged Mortaga, Lukas Rudat, and Luisa Winzer for their support in developing the study environments for the Hanabi, resp. LLM study, and Michelle Wrage for her support in data analysis. Finally, we want to thank Mark Dunn and Silvia Wersing for supporting the scale translations.

This work was supported by the Honda Research Institute Europe GmbH (project CoCharge) and by the first author's DenkRaum Fellowship from the Universities of Lübeck and Kiel (2023–2025).

\section*{Declaration of generative AI and AI-assisted technologies in the writing process}

During the preparation of this work the authors used the large language model Claude (Sonnet 4.6) in order to improve language quality. After using this tool, the authors reviewed and edited the content as needed and take full responsibility for the content of the published article.

\printbibliography

\appendix
\section{Perceived cooperativity scale}\label{app:pcs}

\begin{table}[H]
        \caption{Perceived Cooperativity Scale (Agent Version, PCS-A)}
            \renewcommand{\arraystretch}{1.2}
    \begin{tabular}    {>{\raggedright\arraybackslash}p{0.07\linewidth}>{\raggedright\arraybackslash}p{0.4\linewidth}>{\raggedright\arraybackslash}p{0.46\linewidth}}
    \toprule
    Item \#     & English version  & German version\\
    \midrule
 & I had the feeling that ...&Ich hatte das Gefühl, dass ...\\
 \midrule
     PCS01    &  ... the agent could communicate its status to me. &  ... der Agent mir seinen Status mitteilen konnte.\\
     PCS02    &  ... the agent made clear what it wanted to achieve.& ... der Agent klarmachte, was er erreichen wollte.\\
     PCS03    &  ... the agent kept me informed about its actions.& ... der Agent mich über sein Vorgehen auf dem Laufenden gehalten hat.\\
     PCS04    &  ... the agent could detect difficulties on the way to achieving the goal.& ... der Agent mögliche Schwierigkeiten bei der Zielerreichung erkennen konnte. \\
     PCS05    &  ... the agent knew enough of my goals to collaborate with me.& ... der Agent genug über meine Ziele wusste, um mit mir zusammenzuarbeiten.\\
     PCS06    &   ... the agent could adapt its communication to a specific situation.& ... der Agent seine Kommunikation an eine konkrete Situation anpassen konnte.\\
     PCS07   &  ... the agent was able to take or hand over the initiative by itself.& ...der Agent von sich aus die Initiative ergreifen oder abgeben konnte.\\
     PCS08    &   ... the agent was able to react to external guidance. & ... der Agent auf äußere Vorgaben reagieren konnte.\\
     PCS09    &  ... it was predictable how the agent would react in different situations.& ... es vorhersehbar war, wie der Agent sich in unterschiedlichen Situationen verhalten würde.\\
     PCS10 & ... the agent was dependable. & ... der Agent verlässlich war.\\ 
 PCS11&  ... the way in which the agent proceeded could be influenced by me. & ... die Art und Weise wie der Agent vorging von mir beeinflusst werden konnte. \\
 PCS12& ... the agent was able to incorporate information I shared.& ... der Agent von mir geteilte Informationen berücksichtigen konnte.\\
 PCS13& ... the agent was able to indicate the most important information. & ... der Agent auf die wichtigsten Informationen hinweisen konnte.\\
 PCS14\textsuperscript{1}& ... the agent was able to realize that we had different approaches to achieving the goal. & ... der Agent erkennen konnte, dass wir unterschiedliche Ansätze zur Zielerreichung hatten.\\
 PCS15& ... the agent tried to solve problems that arose during our cooperation. & ... der Agent versuchte, in unserer Zusammenarbeit auftretende Probleme zu lösen.\\
 \bottomrule
 \multicolumn{3}{p{0.93\linewidth}}{\emph{Note.} Item responses are provided on a 6-point Likert scale from 1 (\emph{completely disagree}) to 6 (\emph{completely agree}). \textsuperscript{1} Item excluded from final scale.}\\
    \end{tabular}
    \label{tab:pcs-a}
\end{table}

\begin{table}[H]
        \caption{Perceived Cooperativity Scale (Human Partner Version, PCS-H)}
            \renewcommand{\arraystretch}{1.2}
    \begin{tabular}    {>{\raggedright\arraybackslash}p{0.07\linewidth}>{\raggedright\arraybackslash}p{0.4\linewidth}>{\raggedright\arraybackslash}p{0.46\linewidth}}
    \toprule
    Item \#     & English version  & German version\\
    \midrule
 & I had the feeling that ...&Ich hatte das Gefühl, dass ...\\
 \midrule
     PCS01    &  ... the person could communicate their status to me. &  ... die Person mir ihren Status mitteilen konnte.\\
     PCS02    &  ... the person made clear what they wanted to achieve. & ... die Person klarmachte, was sie erreichen wollte.\\
     PCS03    &  ... the person kept me informed about their actions. & ... die Person mich über ihr Vorgehen auf dem Laufenden gehalten hat.\\
     PCS04    &  ... the person could detect difficulties on the way to achieving the goal. & ... die Person mögliche Schwierigkeiten bei der Zielerreichung erkennen konnte. \\
     PCS05    &  ... the person knew enough of my goals to collaborate with me. & ... die Person genug über meine Ziele wusste, um mit mir zusammenzuarbeiten.\\
     PCS06    &   ... the person could adapt their communication to a specific situation. & ... die Person ihre Kommunikation an eine bestimmte Situation anpassen konnte.\\
     PCS07   & ... the person was able to take or hand over the initiative on their own. & ... die Person von sich aus die Initiative ergreifen oder abgeben konnte.\\
     PCS08    &  ... the person was able to react to external guidance. & ... die Person auf äußere Vorgaben reagieren konnte.\\
     PCS09    &  ... it was predictable how the person would react in different situations. & ... es vorhersehbar war, wie die Person sich in unterschiedlichen Situationen verhalten würde.\\
     PCS10 & ... the person was dependable. & ... die Person verlässlich war.\\ 
 PCS11&  ... the way in which the person proceeded could be influenced by me. & ... die Art und Weise wie die Person vorging von mir beeinflusst werden konnte. \\
 PCS12& ... the person was able to incorporate information I shared. & ... die Person von mir geteilte Informationen berücksichtigen konnte.\\
 PCS13& ...  the person was able to indicate the most important information. & ... die Person auf die wichtigsten Informationen hinweisen konnte.\\
 PCS14\textsuperscript{1}& ... the person was able to realize that we had different approaches to achieving the goal. & ... die Person erkennen konnte, dass wir unterschiedliche Ansätze zur Zielerreichung hatten.\\
 PCS15& ... the person tried to solve problems that arose during our cooperation. & ... die Person versuchte, in unserer Zusammenarbeit auftretende Probleme zu lösen.\\
 \bottomrule
 \multicolumn{3}{p{0.93\linewidth}}{\emph{Note.} Item responses are provided on a 6-point Likert scale from 1 (\emph{completely disagree}) to 6 (\emph{completely agree}). \textsuperscript{1} Item excluded from final scale.}\\
    \end{tabular}
    \label{tab:pcs-h}
\end{table}

\section{Teaming perception scale}\label{app:tps}

\begin{table}[H]
        \caption{Teaming Perception Scale (Agent Version, TPS-A)}
            \renewcommand{\arraystretch}{1.2}
    \begin{tabular}    {>{\raggedright\arraybackslash}p{0.07\linewidth}>{\raggedright\arraybackslash}p{0.4\linewidth}>{\raggedright\arraybackslash}p{0.46\linewidth}}
    \toprule
    Item \#     & English version  & German version\\
    \midrule
     TPS01    &  The agent and I were a team. &  Der Agent und ich waren ein Team.\\
     TPS02    &  The agent and I worked together towards a goal. & Der Agent und ich haben gemeinsam auf ein Ziel hingearbeitet.\\
    TPS03    &  The agent and I supported each other. & Der Agent und ich haben uns gegenseitig unterstützt.\\
     TPS04    &  I contributed to achieving the common goal. & Ich habe zum Erreichen des gemeinsamen Ziels beigetragen. \\
     TPS05    &  I put aside my own goals in favor of the agent. & Ich habe meine eigenen Ziele zugunsten des Agenten zurückgestellt.\\
     TPS06    &   I supported the agent. & Ich habe den Agenten unterstützt.\\
     TPS07   &  The agent contributed to achieving the common goal. & Der Agent hat zum Erreichen des gemeinsamen Ziels beigetragen.\\
     TPS08    &   The agent put aside its goals in favor of me.  & Der Agent hat seine eigenen Ziele zu meinen Gunsten zurückgestellt.\\
     TPS09    &  The agent supported me. & Der Agent hat mich unterstützt.\\
 \bottomrule
 \multicolumn{3}{l}{\emph{Note.} Item responses are provided on a 6-point Likert scale from 1 (\emph{completely disagree}) to 6 (\emph{completely agree}).}\\
    \end{tabular}
    \label{tab:tps-a}
\end{table}

\begin{table}[H]
        \caption{Teaming Perception Scale (Human Partner Version, TPS-H)}
            \renewcommand{\arraystretch}{1.2}
    \begin{tabular}    {>{\raggedright\arraybackslash}p{0.07\linewidth}>{\raggedright\arraybackslash}p{0.4\linewidth}>{\raggedright\arraybackslash}p{0.46\linewidth}}
    \toprule
    Item \#     & English version  & German version\\
    \midrule
     TPS01    &  The person and I were a team. &  Die Person und ich waren ein Team.\\
     TPS02    &  The person and I worked together towards a goal. & Die Person und ich haben gemeinsam auf ein Ziel hingearbeitet.\\
    TPS03    &  The person and I supported each other. & Die Person und ich haben uns gegenseitig unterstützt.\\
     TPS04    &  I contributed to achieving the common goal. & Ich habe zum Erreichen des gemeinsamen Ziels beigetragen. \\
     TPS05    &  I put aside my own goals in favor of the person. & Ich habe meine eigenen Ziele zugunsten der Person zurückgestellt.\\
     TPS06    &   I supported the person. & Ich habe die Person unterstützt.\\
     TPS07   &  The person contributed to achieving the common goal. & Die Person hat zum Erreichen des gemeinsamen Ziels beigetragen.\\
     TPS08    &   The person put aside their goals in favor of me.  & Die Person hat ihre eigenen Ziele zu meinen Gunsten zurückgestellt.\\
     TPS09    &  The person supported me. & Die Person hat mich unterstützt.\\
 \bottomrule
 \multicolumn{3}{l}{\emph{Note.} Item responses are provided on a 6-point Likert scale from 1 (\emph{completely disagree}) to 6 (\emph{completely agree}).}\\
    \end{tabular}
    \label{tab:tps-h}
\end{table}

\end{document}